\documentclass[11pt,a4paper,reprint,notitlepage,superscriptaddress,aps,pra]{revtex4-2}

\usepackage{amsmath,amssymb,amsfonts} % standard AMS packages
\usepackage{physics}

\usepackage{microtype}

\usepackage[svgnames,dvipsnames]{xcolor}
\usepackage{graphicx}
\definecolor{NewBlue}{rgb}{0.1, 0.1, 0.7}
\definecolor{NewRed}{rgb}{0.7, 0.1, 0.1}
\usepackage[colorlinks,
    linkcolor=Maroon,
    citecolor=NewBlue,
    urlcolor=ForestGreen]{hyperref}

\usepackage[capitalise]{cleveref}

\usepackage[
    exponent-product=\cdot,
    range-phrase=\text{--},
    range-units=single,
    separate-uncertainty,
]{siunitx} 

\newcommand{\change}[1]{#1}

\renewcommand{\t}[1]{\mathrm{{#1}}}

\newcommand{\LigoMIT}{LIGO Laboratory, Massachusetts Institute of Technology, Cambridge, MA 02139}
\newcommand{\MechMIT}{Department of Mechanical Engineering, Massachusetts Institute of Technology, Cambridge, MA 02139}
\newcommand{\PhysMIT}{Department of Physics, Massachusetts Institute of Technology, Cambridge, MA 02139}
\newcommand{\MKI}{MIT Kavli Institute for Astrophysics and Space Research, Cambridge, MA 02139}

\begin{document}

\title{Scheme for continuous force detection with a single electron at the level of $10^{-27}\ \t{N}$}
\author{Dominika \v{D}urov\v{c}\'{i}kov\'{a}}
\email{dominika@mit.edu}
\affiliation{\PhysMIT}
\affiliation{\MKI}
\author{Vivishek Sudhir}
\email{vivishek@mit.edu}
\affiliation{\LigoMIT}
\affiliation{\MechMIT}
\date{\today}

% \begin{abstract}
%   We propose and analyze in detail a new scheme for high-sensitivity continuous force detection using a single trapped electron transducer. Force detection is enabled through continuous measurement of the electron's displacement by coupling the motion of the trapped electron to a microwave cavity field via image currents 
%   induced in an antenna. We derive the fundamental and technical limits to the sensitivity with which the electron's motion can be monitored. Specifically, we show that despite the disparity in size between that of a single electron and the wavelength of the microwave field, it is possible in principle to continuously monitor the charge's zero-point motion and use it as a force detector with a sensitivity as low as $6\times 10^{-27} \t{\ N}/ \sqrt{\t{Hz}}$ in the gigahertz regime. This sensitivity improves on the state-of-the-art by four orders of magnitude and thus paves the way to novel precision experiments at the interface of quantum physics and gravity.
% \end{abstract}

\begin{abstract}
The detection of weak forces is a central problem in physics and engineering, ranging in importance from
fundamental pursuits such as precision tests of gravity, gravitational-wave detection, and
searches for dark matter, to applications such as force microscopy. These pursuits require a low-mass mechanical
force transducer with a high quality factor, whose motion can be measured in a quantum-noise-limited manner. 
Here we study the ultimate example of such a transducer: a single
trapped electron. We propose and analyze in detail a new scheme for high-sensitivity continuous force detection using a single trapped electron 
whose motion is coupled to a microwave cavity field via image currents 
induced in an antenna. We derive the fundamental and technical limits to the sensitivity of this scheme 
and show that despite the disparity in size between that of a single electron and the wavelength of the microwave field, it is possible 
to continuously monitor the charge's zero-point motion and use it as a force detector with a sensitivity as low 
as \change{$6\times 10^{-27}\, \t{N}/ \sqrt{\t{Hz}}$} 
in the gigahertz regime. 
This sensitivity improves on the state-of-the-art by four orders of magnitude and thus paves the way to novel precision experiments.
\end{abstract}

\maketitle

\section{Introduction}

High-precision time-continuous force sensing has been a tool of fundamental physics for over two centuries,
ranging from precision tests of gravity \cite{Cav,Dicke,Adel09,Wag12,SpeQui14} and gravitational-wave 
detection \cite{adhikari_gravitational_2014,buikema_sensitivity_2020}, 
to recent advances in macroscopic quantum 
mechanics \cite{SchRouk05,aspelmeyer_cavity_2014,whittle_approaching_2021}, 
and future avenues for dark matter detection \cite{carney_mechanical_2021,Carn21,BudUlm22}.
At the heart of such measurements is a mechanical oscillator, whose motion is monitored 
for its response to an external force of interest. 

\begin{figure}[h!]
    \centering
    \includegraphics[width=0.9\columnwidth]{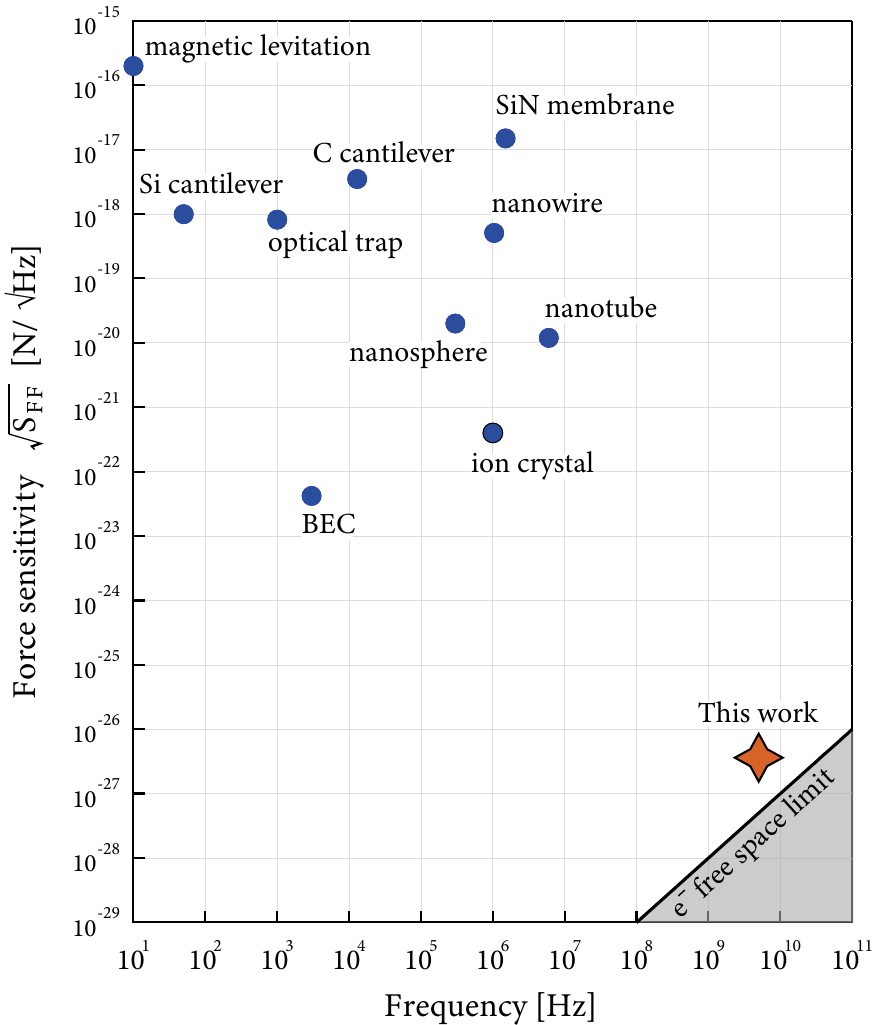}
    \caption{A survey of force sensitivities measured in a variety of platforms: 
    Bose-Einstein condensate (BEC) \cite{schreppler_optically_2014},
    trapped ion crystals \cite{biercuk_ultrasensitive_2010}, optical traps \cite{monteiro_force_2020,delic_cooling_2020}, magnetically levitated spheres \cite{timberlake_acceleration_2019,lewandowski_high-sensitivity_2021}, nanomechanical oscillators \cite{teufel_nanomechanical_2009,gavartin_hybrid_2012,moser_ultrasensitive_2013}, and micro-cantilevers \cite{mamin_sub-attonewton_2001,tao_single-crystal_2014}. The force sensitivity that is feasible with our proposed setup is shown as an orange star. The shaded region shows the fundamental limit to the achievable sensitivity of a single-electron transducer based on its thermal and zero-point motion (in free space at a temperature of 4K).}
    \label{fig:force_survey}
\end{figure}

The fundamental limit to the sensitivity of such a force transducer is set by
its intrinsic motion, due to noise from fluctuating thermal (``Brownian'') forces and its
zero-point motion. The spectral density of this intrinsic force noise is given
by the fluctuation-dissipation theorem (FDT)
\cite{callen_irreversibility_1951,kubo_fluctuation-dissipation_1966}:
\begin{equation}\label{eq:SFFint}
    \bar{S}_{FF}^\t{int}[\Omega] = 2 \hbar m \Omega \gamma \left( n_\t{th}[\Omega]+\tfrac{1}{2} \right),
\end{equation}
where for an oscillator of mass $m$, $\gamma$ is its damping rate, $n_\t{th}
[\Omega] = (e^{\hbar \Omega/k_B T}-1)^{-1}$ is its average thermal occupation,
and the $1/2$ represents its quantum mechanical zero-point motion. The highest
sensitivies to external forces are achieved when this intrinsic force noise is
minimized. According to \cref{eq:SFFint}, a small-mass, low-loss mechanical
oscillator operated at low temperature ($T < \hbar \Omega/k_B$) is the right
strategy. \change{(Note that this contrasts with displacement sensing, such as in gravitational-wave
detectors \cite{adhikari_gravitational_2014}, where the sensitivity scales inversely with the oscillator mass.)}
Indeed, the dogged pursuit of \change{an increase in force sensitivity} has driven the field of
nanomechanics \cite{newell_miniaturization_1968,SidYan95,
ekinci_nanoelectromechanical_2005,bachtold_mesoscopic_2022}, culminating in a
demonstrated force sensitivity of \cite{moser_ultrasensitive_2013}
$10^{-20}\t{N/\sqrt{Hz}}$ using a carbon nanotube operated at $\SI{1.2}{K}$.

It is worth asking what force sensitivity can be achieved given contemporary
understanding of science and technology. A single trapped electron is the
lowest-mass object that can be trapped to form a harmonic oscillator. In a
Penning trap, they can be trapped with high motional frequencies ($\Omega \sim
2\pi \cdot \SI{10}{GHz}$), and feature exceptionally low decoherence
rates \cite{borchert_measurement_2019} $\gamma n_\t{th} \sim 2\pi\cdot
10^{-3}\, \t{Hz}$, which indicate the possibility of achieving intrinsic force
noise as low as $\bar{S}_{FF}^\t{int} \sim 10^{-27}\, \t{N/\sqrt{Hz}}$ ---
about four orders of magnitude smaller than the state-of-the-art (see
\cref{fig:force_survey}).

The challenge is to perform continuous measurement of the motion of a single
electron such that the sensitivity of the measurement reaches the limit due to
its intrinsic motion. 
The primary obstruction is that individual sub-/atomic
particles, due to their small size (compared to electromagnetic wavelength), do
not strongly interact with electromagnetic fields for their motion to be
monitored with the requisite precision
\cite{stokes_precision_1991,thomas_precision_1995,HoodKimb00,schreppler_optically_2014}.

We propose a novel scheme for the continuous measurement of motion
of a single electron in a Penning trap that is expected to achieve a force
sensitivity as low as \change{$6\times 10^{-27}\, \t{N}/ \sqrt{\t{Hz}}$} . Our scheme consists
of coupling the trapped electron to microwave fields via an antenna, which
magnifies the dipole moment of the sub-atomic electron to macroscopic scales,
thus enabling its coupling to the microcave cavity field. \change{For the first time, we} derive the quantum
theory of this novel interaction and use it to account for the inevitable
quantum-mechanical tradeoff between measurement imprecision and backaction
noises for \change{continuous force sensing using a microwave-coupled single electron in a Penning trap.
(This objective, and its requirements, are qualitatively different from existing single-particle Penning trap
experiments \cite{blaum06,blaum_penning_2010} to precisely measure motional frequencies or for mass spectrometry.)}
We summarize the different technical noise sources that are
expected to limit the practical performance of such a setup. The resulting
force sensitivity over a few GHz bandwidth improves on the state-of-the-art by
four orders of magnitude and thus promises to enable new experiments at the
interface of quantum physics and gravity \cite{Soda22} and accelerate searches for dark
matter candidates \cite{carney_mechanical_2021,Carn21,BudUlm22}.

\section{An electron strongly coupled to a microwave cavity field} % ~2580 words

% This section outlines the details of the proposed measurement scheme as well as the relevant readout and technical noise sources that limit its force sensitivity.

The lowest-order interaction between a charged particle and the electromagnetic field is of the electric dipole form,
which however is negligible if the spatial extent of the dipole moment of the particle is much smaller compared
to the wavelength of the field. In what follows we will show that antenna-mediated coupling of the motion of a
trapped electron to a microwave cavity field forms a crucial piece in bridging the scale-gap between a single 
electron and microwave fields. That is because the antenna magnifies the dipole moment of a single charge. 
In \cref{sec:coupling}, we derive the quantum theory of this antenna-mediated interaction.

The coupling \change{is mediated by the} image current induced on the antenna by the oscillating electron. These image 
charges have been shown to not only transduce the motion of the trapped particle, but also act back and 
affect its motion \cite{sturm_phase-sensitive_2011,van_dyck_number_1989,porto_series_2001}. 
Fundamentally, the measurement sensitivity and back-action must be limited by quantum fluctuations inherent
in the measurement process. In \cref{sec:noiseMeas}, we use our quantum theory of the interaction to derive,
for the first time, the quantum mechanical limits to continuous displacement measurement of a trapped charged particle
via its coupling to microwave fields. Finally, in \cref{sec:noise}, we consider relevant technical noises
that can limit sensitivity and produce excess classical back-action.

\subsection{Quantum theory of electron-microwave coupling}\label{sec:coupling}

\begin{figure}
    \centering
    \includegraphics[width=\linewidth]{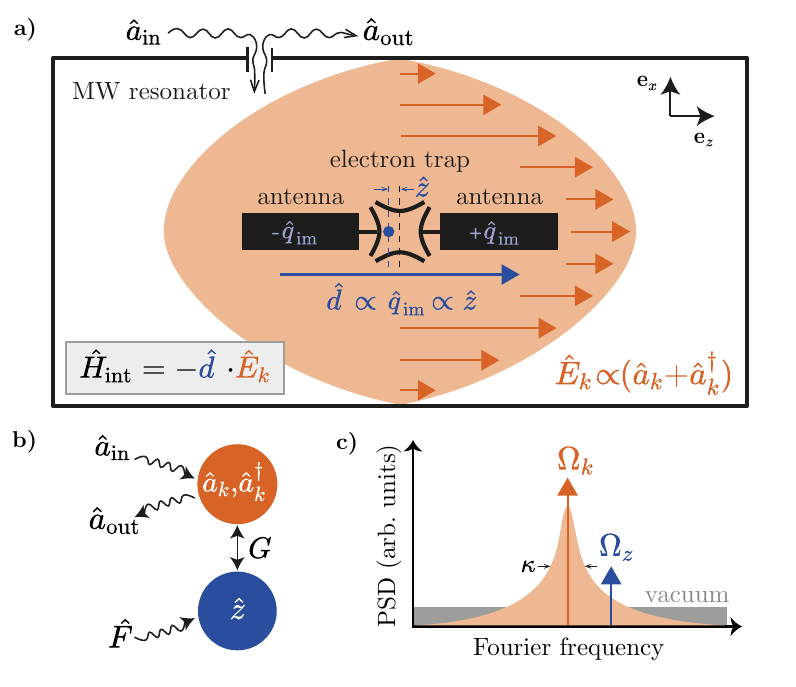} % ~190 words
    \caption{a) Schematic of the proposed coupling scheme. The axial motion, $\hat{z}$ of the electron in the Penning trap is transduced to position-dependent image charges $\pm \hat{q}_\t{im}$ in the antenna. This amplifies the effective dipole moment of the electron oscillations $\hat{d}$, in turn strengthening the coupling between the microwave field in the resonator, $\hat{E}_k$, and the electron displacement, $\hat{z}$. b) Diagram of the measurement scheme in the quantum Langevin description. The microwave cavity modes, $\hat{a}_k$ and $\hat{a}_k^\dagger$, can be in general pumped with external microwave drive $\hat{a}_\t{in}$. The cavity modes then couple to the axial mode of the electron, $\hat{z}$ that is subject to a force that we wish to measure, $\hat{F}$. The output field $\hat{a}_\t{out}$ is read out, containing information about the electron motion and the force $\hat{F}$. c) Frequency picture of this coupling scheme. The axial motional frequency, $\Omega_z$, has to be off resonance with the cavity frequency, $\Omega_k$ (as resonance amplifies dynamical backaction and thus amplifies noise), but not much further away from the resonance than the cavity linewidth (as the electron and the antenna need to be able to radiate into the cavity or imprecision is increased). The tradeoff between these two regimes defines the sensitivity band of this detector.}
    \label{fig:schematic}
\end{figure}

Consider the scheme depicted in \cref{fig:schematic}a: a single electron (mass $m$, charge $e$) is 
trapped in a Penning trap \citep{brown_geonium_1986} 
whose electrodes extend to an antenna embedded in a microwave cavity. The electron 
oscillates in the axial ($\textbf{e}_z$) direction between two trap electrodes separated by a distance $2z_0$.
This motion ($\hat{z}$) induces image charges on the electrodes \cite{wineland_principles_1975}
\begin{equation}
    \pm q_\t{im}(t) = \pm \alpha e \frac{z(t)}{2z_0},
\end{equation}
where $\alpha$ is a factor of the order unity that depends on the electrode geometry. For the sake of simplicity, $\alpha=1$ is the approximation that we will use here. The creation of an image charge at the trap electrode is due to redistribution of electrons in the conducting antenna. Because of the symmetry of the double-sided, linear antenna, this redistribution shows up as a surplus of charge on one side of the antenna and a deficit of charge on the other side. The induced image charges at the ends of the antenna in turn magnify the dipole moment of the sub-atomic electron to macroscopic scales; 
the enlarged dipole moment of the antenna couples to a microwave field mode ($\hat{a}_k$) 
stored in the surrounding cavity, whose leakage ($\hat{a}_\t{out}$) is measured.
The antenna thus forms a crucial piece in overcoming the size discrepancy between the electron 
and the cm-scale microwaves. 

In order to describe the interaction between the electron and the intracavity microwave field, we model the resultant image current in the antenna as a collection of moving electrons that interact with the cavity electromagnetic field just outside the antenna. The Hamiltonian of such an interaction can be derived starting from the Lagrangian of non-relativistic electrodynamics (\cite{cohen-tannoudji_photons_2004} II.B.1a)
\begin{equation}\label{eq:LEM}
    \begin{split}
        \mathcal{L} = & \sum_i \frac{m}{2} \dot{\textbf{x}}_i^2 + \int d^3\textbf{x} \left( \frac{\epsilon_0}{2}\textbf{E}^2 - \frac{1}{2\mu_0}\textbf{B}^2\right) \\
        & + \sum_i \left( e \dot{\textbf{x}}_i \cdot \textbf{A}(\textbf{x}_i) - e U(\textbf{x}_i)\right),
    \end{split} 
\end{equation}
where $\textbf{x}_i,\dot{\textbf{x}}_i$ are the displacements and velocities of the moving charges in the antenna, 
and $U(\textbf{x}_i)$ and $\textbf{A}(\textbf{x}_i)$ are the position dependent scalar and vector potentials 
of the intracavity field. 

In order to arrive at an interaction Hamiltonian expressed in terms of the electric and magnetic fields, 
we will perform a canonical transformation of $\mathcal{L}$. As a more convenient representation of the positions and velocities of the collection of electrons, we define the generalized polarization and magnetization, respectively, \cite{vukics_gauge-invariant_2021,cohen-tannoudji_photons_2004}
\begin{equation}\label{eq:P}
    \textbf{P}(\textbf{x},t) = \sum_i e \textbf{x}_i(t) \int ds \delta(\textbf{x}-s\textbf{x}_i(t)),
\end{equation}
\begin{equation}\label{eq:M}
    \textbf{M}(\textbf{x},t) = \sum_i e \textbf{x}_i(t) \times \dot{\textbf{x}}_i(t) \int ds \delta(\textbf{x}-s\textbf{x}_i(t)),
\end{equation}
where $s$ is a dummy parameter that sums over a collection if infinitesimal dipoles that make up the effective dipole moment between the point of interest, $\textbf{x}$, and a charge $\textbf{x}_i$ (refer to \cite{cohen-tannoudji_photons_2004} IV.C.1a). Note that the electric current density in the antenna can have contributions both from a time-varying polarization as well as a curl in the magnetization, 
\begin{equation}\label{eq:Jfull}
    \textbf{J}(\textbf{x},t) = \dot{\textbf{P}} + \nabla\times\textbf{M}.
\end{equation}
We can now rewrite \cref{eq:LEM} using the definitions in \cref{eq:P,eq:M} and dropping all terms which do not contribute to the dynamics,
\begin{equation}
    \begin{split}
        \mathcal{L} = & \sum_i \frac{m}{2} \dot{\textbf{x}}_i^2 + \int d^3\textbf{x} \left( \frac{\epsilon_0}{2}\textbf{E}_\perp^2 - \frac{1}{2\mu_0}\textbf{B}^2\right) \\
        & + \int d^3\textbf{x} \left( \dot{\textbf{P}}_\perp + \nabla\times\textbf{M} \right)\cdot\textbf{A}_\perp.
    \end{split} 
\end{equation}
where the transverse quantities $\vb{E}_\perp,\vb{P}_\perp, \vb{A}_\perp$ are defined as $\nabla \cdot \vb{E}_\perp = 0$ et cetera.

The canonical transformation of interest is of the Power-Zienau-Woolley (PZW) 
type \cite{Bab83,vukics_fundamental_2015,vukics_gauge-invariant_2021,cohen-tannoudji_photons_2004}, which can be
implemented at the level of the Lagrangian by adding a total time derivative (which obviously does not contribute to
the dynamics, since the action is invariant to addition of a total time derivative):
\begin{equation}\label{eq:PZWlagrangian}
\begin{split}
	\mathcal{L}_\t{PZW} &= \mathcal{L} - \frac{d}{dt} \int d^3\textbf{x} \textbf{P}_\perp \cdot \textbf{A}_\perp \\
	&= \sum_i \frac{m}{2} \dot{\textbf{x}}_i^2 + \int d\textbf{x}^3 \left( \frac{\epsilon_0}{2}\textbf{E}_\perp^2 
		- \frac{1}{2\mu_0}\textbf{B}^2\right) \\
    &\qquad\qquad +\int d\textbf{x}^3 \left( \textbf{P}_\perp\cdot\textbf{E}_\perp + \textbf{M}\cdot\textbf{B} \right).
\end{split}
\end{equation}
Here we have made use of the fact that $\int d^3\textbf{x}(\nabla\times\textbf{M})\cdot\textbf{A}_\perp = \int d^3\textbf{x}\, \textbf{M} \cdot (\nabla\times\textbf{A}_\perp)$.
Defining the conjugate momenta for the particles and the field,
\begin{equation}
    \textbf{p}_i = \frac{\partial \mathcal{L}_\t{PZW}}{\partial \dot{\textbf{x}}_i} = m\dot{\textbf{x}}_i - e\textbf{x}_i \times \int ds \textbf{B} (s\textbf{x}_i) ,
\end{equation}
\begin{equation}
    \boldsymbol{\pi} = \frac{\partial \mathcal{L}_\t{PZW}}{\partial \dot{\textbf{A}}} = - \epsilon_0 \textbf{E}_\perp - \textbf{P}_\perp,
\end{equation}
and performing a Legendre transformation gives the Hamiltonian
\begin{equation}\label{eq:HPZW}
    \begin{split}
        \mathcal{H}_\t{PZW} = & \sum_i \frac{1}{2m} \left(\textbf{p}_i + \frac{e}{m} \textbf{x}_i \times \int ds \textbf{B} (s\textbf{x}_i)  \right)^2 \\
        &  + \int d\textbf{x}^3 \left( \frac{1}{2\epsilon_0}\textbf{D}_\perp^2 + \frac{1}{2\mu_0}\textbf{B}^2\right) \\
        & - \int d\textbf{x}^3 \textbf{P}_\perp \cdot\textbf{E}_\perp - \int d\textbf{x}^3 \frac{1}{2\epsilon_0}\textbf{P}_\perp^2.
    \end{split}
\end{equation}
Here, $\textbf{D}_\perp \equiv \epsilon_0 \textbf{E}_\perp + \textbf{P}_\perp$ is the displacement field. 
The first line of \cref{eq:HPZW} describes the charges in the antenna (note that the canonical momentum is now dependent on the magnetic field), the second line describes the electromagnetic field in and around the antenna, 
and the third line contains the charge-field coupling term of our interest and a self-interaction term. 
Note that the term $\int d^3\textbf{x}\textbf{M}\cdot\textbf{B}$ in \cref{eq:PZWlagrangian} has been cancelled by the term $\sum_i \dot{\textbf{x}}_i \cdot \left(- e\textbf{x}_i \times \int ds \textbf{B} (s\textbf{x}_i) \right) = \sum_i e\textbf{x}_i \times \dot{\textbf{x}}_i \int ds \textbf{B} (s\textbf{x}_i)$.
The self-interaction term can be treated as an additional scalar potential term as long as we can approximate the 
charge ensemble as noninteracting; i.e. the only interaction between the charges in the antenna is via their interaction with the external electromagnetic field \cite{vukics_fundamental_2015}. 
In the Coulomb gauge, scalar potentials do not lead to dynamics, and so this term can be dropped. 

Thus, \cref{eq:HPZW} shows that the interaction of the antenna current with the electromagnetic field in the cavity is twofold: first, the canonical momentum is modified by a Lorentz force term (first line of \cref{eq:HPZW}), and second, we recover the general form of the dipole-like interaction Hamiltonian 
\begin{equation}\label{eq:Hintgeneral}
    H_\t{int} = - \int d\textbf{x}^3 \textbf{P}_\perp \cdot\textbf{E}_\perp.
\end{equation}
Here, both $\textbf{P}_\perp$ and $\textbf{E}_\perp$ depends on both the spatial coordinate and time.
Importantly, $\vb{P}_\perp$ is the generalized polarization of the antenna.

In this manuscript, we consider the simplest scenario of a symmetric, double-sided, linear antenna (total length $l$) coupled to a TE mode of the cavity. To derive a simpler form of the interaction Hamiltonian in \cref{eq:Hintgeneral}, we can Taylor expand the intracavity electric field around the center of the antenna, $\textbf{E}_\perp (\textbf{x},t)|_\textbf{0} = \textbf{E}_\perp (\textbf{0},t) + \textbf{x}\cdot (\nabla \cdot \textbf{E}_\perp (\textbf{x},t))|_\textbf{0} + \mathcal{O}(\abs{\textbf{x}}^2)$, and use the definition of the polarization in \cref{eq:P} to integrate over the 
$\delta$-functions: $\int_0^1 ds \int d^3\textbf{x}\delta(\textbf{x}-s\textbf{x}_i) = 1$, 
$\int_0^1 ds \int d^3\textbf{x}\delta(\textbf{x}-s\textbf{x}_i)\cdot\textbf{x} = \textbf{x}_i$, to yield
% \begin{equation}
%     \int_0^1 ds \int d^3\textbf{x}\delta(\textbf{x}-s\textbf{x}_i) = 1,
% \end{equation}
% \begin{equation}
%     \int_0^1 ds \int d^3\textbf{x}\delta(\textbf{x}-s\textbf{x}_i)\cdot\textbf{x} = \textbf{x}_i.
% \end{equation}
% This procedure yields
\begin{equation}\label{eq:HintTaylor}
    \begin{split}
        H_\t{int} = & - \sum_i e \textbf{x}_i\cdot\textbf{E}_\perp(\textbf{0},t)\\ 
        & - \sum_i e \abs{\textbf{x}_i}^2 \left( \nabla \cdot \textbf{E}_\perp (\textbf{x},t) \right) |_\textbf{0}
        + \mathcal{O}(\abs{\textbf{x}_i}^3),
    \end{split}
\end{equation}
where the sum is over the antenna charges.

\begin{figure}[t!]
    \centering
    \includegraphics[width=\linewidth]{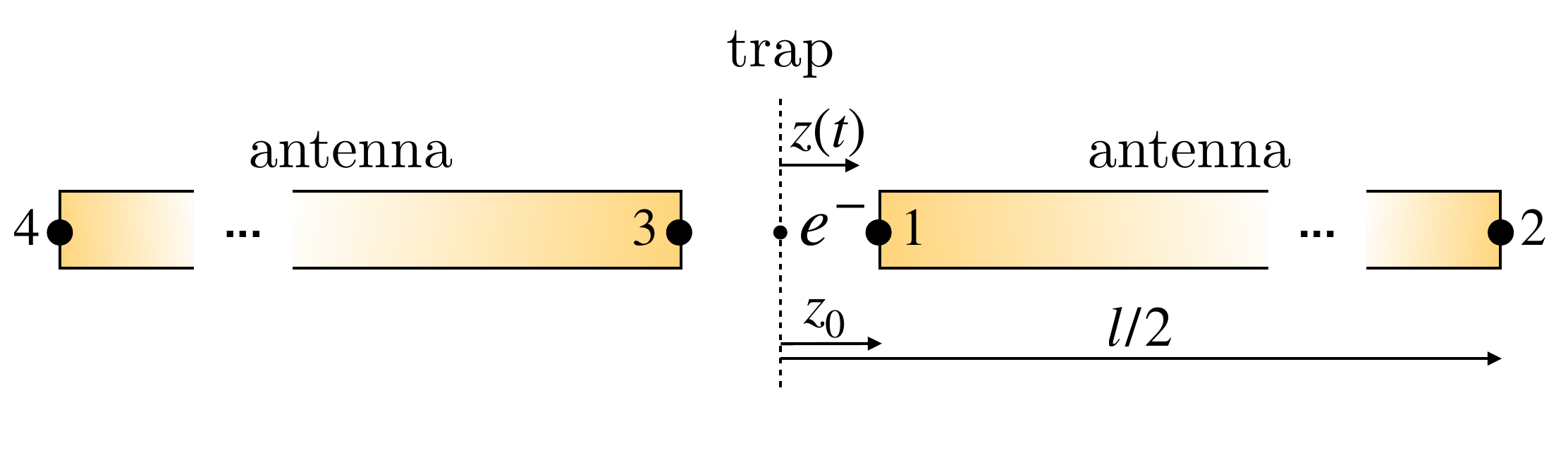}
    \caption{A schematic of the double-sided antenna geometry around the electron trap. The sum over charges in the antenna in \cref{eq:HintTaylor} can be cast in terms of a sum over image charges at the four end points of the sectioned antenna, labelled $1$-$4$.}
    \label{fig:geometry}
\end{figure}

Finally, we can relate the collection of microscopic charges in the antenna to the image charge induced on it
by the oscillating electron as follows.
The sum over antenna charges labelled by the subscript $i$ can be cast in terms of a sum over four charges corresponding to the end points of the two sections of the linear antenna (one section on each side of the trap, see \cref{fig:geometry}), i.e.
\begin{equation}
    \abs{\textbf{x}_1} = \abs{\textbf{x}_3} = z_0,
\end{equation}
\begin{equation}
    \textbf{x}_1 = -\textbf{x}_3
\end{equation}
for the image charges at the trap electrodes, and 
\begin{equation}
    \abs{\textbf{x}_2} = \abs{\textbf{x}_4} = \frac{l}{2},
\end{equation}
\begin{equation}
    \textbf{x}_2 = -\textbf{x}_4
\end{equation}
for the image charges at the far end of both antenna sections. We can do this decomposition because the creation of an image charge at the trap electrode is due to redistribution of electrons in the conducting antenna. Because of this charge redistribution, the two ends of an antenna section need to carry an opposite image charge.

The image charge on the surrounding conductors is created as soon as the trapped electron is present -- we label this ``stationary'' charge on each of the two axial electrodes as $\bar{q}_\t{im} = \alpha e/2$. Here, $\alpha\leq 1$ is a geometric factor that accounts for the fact that some of the field lines of the trapped electron may leak out and not contribute to the creation of an axial image charge (and so the total trap image charge need not be equal to $e$). As the trapped electron moves around in the trap, it induces an additional image charge correction $q_\t{im}$ \cite{wineland_principles_1975},
\begin{equation}
    q_\t{im}(t) = \alpha e \frac{z(t)}{2z_0},
\end{equation}
where the role of $\alpha$ is identical to the stationary image charge case. Because of the symmetry of the double-sided, linear antenna, this correction shows up as a surplus of charge on one side of the trap and a deficit of charge on the other side. 
\change{However, due to self-capacitance effects of the electrode and the antenna, the image charge on the far side of both antenna sections will be reduced by a factor $\xi$ given by \cite{van_horne_improved_2024}
\begin{equation}
    \xi = \frac{C_{\rm antenna}}{C_{\rm total}},
\end{equation}
where we model the total self-capacitance as a sum of the self-capacitances of the antenna and the electrode, $C_{\rm total} = C_{\rm antenna} + C_{\rm electrode}$ (and neglect any short connecting wires as their self-capacitance will be negligible).}
Hence, we can rewrite the sum as
\begin{equation}
    \begin{split}
        \sum_i e \textbf{x}_i & = \left( \bar{q}_\t{im} + q_\t{im}(t) \right) \textbf{x}_1 + \left( - \bar{q}_\t{im} - q_\t{im}(t) \right) \xi \textbf{x}_2 \\
        &  + \left( \bar{q}_\t{im} - q_\t{im}(t) \right) \textbf{x}_3 + \left( - \bar{q}_\t{im} + q_\t{im}(t) \right) \xi \textbf{x}_4 \\
        & = 2 q_\t{im}(t) \left( \textbf{x}_1 - \xi \textbf{x}_2 \right) \\
        & \approx - 2 \xi  q_\t{im}(t) \textbf{x}_2.
    \end{split}
\end{equation}
Here, the last line follows as $\abs{\textbf{x}_1} \ll \abs{\textbf{x}_2} = l/2$. 

The first-order term of the interaction Hamiltonian \change{thus} simplifies to 
\begin{equation}\label{eq:dipoleH}
        H_\t{int}^{(1)} = - \left( \xi \alpha e \frac{l}{2z_0} \textbf{z}(t) \right) \cdot\textbf{E}(\textbf{0},t) = - \textbf{d}(t) \cdot\textbf{E}_\perp(\textbf{0},t).
\end{equation}
\cref{eq:dipoleH} is the well-known electric dipole coupling, where the dipole moment corresponding to the antenna is given by the bracketed term in the first equality. We can further express the electric field in terms of the cavity modes labelled by $k$,
\begin{equation}
    H_\t{int}^{(1)} = - \textbf{d}\cdot \sum_k \textbf{E}_k (\textbf{0},t).
\end{equation}
For completeness, we can apply the same procedure to rewrite the second-order correction to the electric dipole Hamiltonian (second term in \cref{eq:HintTaylor}) as 
\begin{equation}
    H_\t{int}^{(2)} = - \alpha e \left( \frac{l}{2} \right)^2 \sum_k \left( \nabla \cdot \textbf{E}_k (\textbf{x},t) \right) |_\textbf{0}.
\end{equation}

From the expression in \cref{eq:dipoleH}, we can observe that only the $z$ component of the intracavity microwave field can interact with the induced dipole moment. Quantizing this field yields
\begin{equation}
    \hat{E} = \sum_k \mathcal{E}_{\t{zp},k} (\hat{a}_k + \hat{a}_k^\dagger)
\end{equation}
with $\mathcal{E}_\t{zp} = \sqrt{\frac{\hbar \Omega_k}{2\epsilon_0 V}}$, where $\hat{a}_k$ and $\hat{a}_k^\dagger$ are the field annihilation and creation operators, $\Omega_k$ is the frequency of the microwave mode labelled by $k$, and $V$ is the mode volume. Henceforth, we will only consider a single mode of the microwave field, $\hat{E}_k$, which we justify by noticing that the sensitivity is strongly dependent on the detuning of the electron motional frequency, $\Omega_z$, from the cavity mode frequency, $\Omega_k$.

Writing the electron axial displacement as an operator, $\hat{z} = z_\t{zp} (\hat{a}_z + \hat{a}_z^\dagger)$ with the zero-point displacement $z_\t{zp} = \sqrt{\frac{\hbar}{2 m \Omega_z}}$, the interaction Hamiltonian in \cref{eq:dipoleH} can be written as
\begin{equation}\label{eq:Hintz}
    \hat{H}_\t{int} = -\hbar G \hat{z}(\hat{a}_k + \hat{a}_k^\dagger),
\end{equation}
with the interaction strength
\begin{equation}\label{eq:gz}
    G = \xi \alpha e\frac{l}{2z_0}\sqrt{\frac{\Omega_k}{2\hbar\epsilon_0V}}.
\end{equation}
\Cref{eq:gz} establishes that a small cavity mode volume and a large ``length enhancement factor'', $\frac{l}{2z_0}$, of the dipole moment are two crucial pieces behind the amplification mechanism that underpins this interaction. \change{The strength of this interaction further depends on two scale factors: $\alpha$ being the extent to which the induced image charges on the trap electrodes reflect the trapped charge (which can easily be $\sim 1$ for a cavity-type trap geometry), and $\xi$, which quantifies how well the induced image charges can ``propagate'' from the trap electrodes to the far end of the antenna (which will be evaluated below).}

Crucially, this setup allows us to induce oscillating image charges on trap electrodes and in the antenna, which are the cornerstone of inducing a large dipole moment that is necessary to couple to the intracavity microwave field.

\subsection{Measurement imprecision and back-action noises}\label{sec:noiseMeas}

The interaction described by $\hat{H}_\t{int}$ is analogous to the linearized radiation-pressure 
interaction between electromagnetic radiation stored in a cavity and its mechanically-compliant end-mirror 
\cite{aspelmeyer_cavity_2014}. In particular, when the microwave cavity is excited with an input field, 
$\hat{a}_\t{in}$, and coupled to the trapped electron via $\hat{H}_\t{int}$, the leakage field, 
$\hat{a}_\t{out}$, can be continuously read out in order to measure the electron motion and thus the 
forces acting on the electron (\cref{fig:schematic}b). However, this act of continuous measurement of
the electron motion is fundamentally accompanied by quantum measurement back-action noise that randomly
disturbs the electron. This can be seen at a first glance by noticing that the electron is susceptible to 
the radiation-pressure-like force $\hat{F}_\t{rp} \equiv \dot{\hat{p}}_z = (i/\hbar)[\hat{H}_\t{int},\hat{p}_z] = \hbar 
G (\hat{a}_k + \hat{a}_k^\dagger)$, so that quantum fluctuations in the intracavity field drive the electron.

This physics can be precisely modeled by considering the Hamiltonian of this open particle-cavity system, 
\begin{equation}
    \begin{split}
        \hat{H} & = \hbar\Omega_k\hat{a}^\dagger_k\hat{a}_k +
        \frac{\hat{p}_z^2}{2m} + \frac{m}{2}\Omega_z^2\hat{z}^2 - 
        \hbar G\hat{z}(\hat{a}_k+\hat{a}^\dagger_k) \\
        & +i\hbar \sqrt{\kappa_\t{in}}(\hat{a}^\dagger_k\hat{a}_\t{in}e^{-i\Omega_\t{in}t}-\hat{a}^\dagger_\t{in}e^{i\Omega_\t{in}t}\hat{a}_k),
    \end{split}
\end{equation}
where the first line corresponds to the particle-cavity system Hamiltonian given by the free Hamiltonians for the axial motion of the electron and the intracavity microwave field, and the interaction Hamiltonian from \cref{eq:dipoleH}, and the second line defines the coupling (characterized by $\kappa_\t{in}$) of our cavity to this probe field. Note that all operators are given in the time-domain, and the Hamiltonian is written from the lab frame perspective.

Using the input-output formalism \cite{gardiner_input_1985}, the Hamiltonian $\hat{H}$ yields the following 
quantum Langevin equation for the intracavity field in the lab frame
\begin{equation}
    \dot{\hat{a}}_k = \left( -i\Omega_k - \frac{\kappa}{2} \right) \hat{a}_k + iG\hat{z} + \sqrt{\kappa_\t{in}} \hat{a}_\t{in} e^{-i\Omega_\t{in}t}.
\end{equation}
Note that the input field rotates at the frequency $\Omega_\t{in}$, and we have used a notation that distinguishes the input linewidth $\kappa_\t{in}$ from the total cavity linewidth $\kappa$ that can additionally  include the linewidth associated with the internal cavity losses, $\kappa=\kappa_\t{in}+\kappa_\t{add}$. However, we restrict our calculation to the case of a lossless cavity, i.e. $\kappa=\kappa_\t{in}$, as this assumption can be easily relaxed. Transforming into the Fourier domain gives \footnote{We are using the following Fourier transform convention: $X[\Omega]=\int dt X(t) e^{i\Omega t}$.}
\begin{equation}\label{eq:ak}
    \hat{a}_k [\Omega] = \chi_k [\Omega] \left( i G \hat{z}[\Omega] + \sqrt{\kappa_\t{in}} \hat{a}_\t{in}[\Omega+\Omega_\t{in}] \right),
\end{equation}
where the cavity susceptibility is defined as $\chi_k [\Omega]=(-i(\Omega-\Omega_k)+\frac{\kappa}{2})^{-1}$.

The equation of motion for the displacement of the electron is given by 
\begin{equation}\label{eq:z}
    \hat{z}[\Omega] =\chi_z[\Omega] \left( \hat{F}_\t{int}[\Omega] + \hat{F}_\t{rp}[\Omega] + \hat{F}_\t{dir}[\Omega] \right),
\end{equation}
where the mechanical susceptibility is defined as $\chi_z[\Omega]=(\Omega_z^2 -\Omega^2 - i\gamma\Omega)^{-1}/m$, $\hat{F}_\t{int}$ is the intrinsic force due to thermal and quantum zero-point motion \cref{eq:SFFint}, $\hat{F}_\t{dir}$ represents all the direct forces acting on the electron, and the radiation pressure force 
is $\hat{F}_\t{rp} = \hbar G (\hat{a}_k+\hat{a}^\dagger_k)$. 

Physically, we expect two kinds of radiation pressure force contributions. One arising from the motion of the electron
itself: the oscillating electron induces a current in the antenna, which radiates into the cavity, gets re-absorbed
by the antenna, and acts back on the electron with a delay. This effect is called dynamical 
backaction \cite{AspMey12}, and it effectively modifies the motional 
frequency of the electron and its damping rate \cite{BrowTan85}. Another contribution is due to the stochastic quantum 
(or classical) fluctuations in the intracavity field, and is independent of the electron motion; this is the stochastic
back-action force. Both can be derived by solving \cref{eq:ak,eq:z} and separating the force term proportional
to the electron motion ($\hat{z}$) and that which is not. This results in the modified equation of motion
of the charge
\begin{equation}\label{eq:z}
    \hat{z}[\Omega] =\chi_z^\t{eff}[\Omega] \left( \hat{F}_\t{int}[\Omega] + \hat{F}_\t{ba}[\Omega] + \hat{F}_\t{dir}[\Omega] \right),
\end{equation}
where $\chi_z^\t{eff}$ is the modified susceptibility due to dynamical back-action, and
$\hat{F}_\t{ba}$ is the stochastic back-action force.

The effective susceptibility due to dynamical back-action is
\begin{equation}
    \chi_z^\t{eff}[\Omega]^{-1} =\chi_z[\Omega]^{-1} + \chi_z^\t{dyn}[\Omega]^{-1},
\end{equation}
with
\begin{equation}
    \begin{split}
        \chi_z^\t{dyn}[\Omega]^{-1} & = - i \hbar G^2 \left( \chi_k[\Omega] - \chi_k^*[\Omega] \right) \\
        & \equiv m\Omega\left( 2 \Omega_\t{ba}[\Omega] - i \gamma_\t{ba}[\Omega] \right).
    \end{split}
\end{equation}
For a high-Q mechanical oscillator, the effect of the dynamical back-action can be approximated as a shift in the 
mechanical frequency and damping, $\Omega_z^\t{eff} = \Omega_z + \Omega_\t{ba}[\Omega_z]$ and 
$\gamma_\t{eff} = \gamma + \gamma_\t{ba}[\Omega_z]$, given by
\begin{equation}
    \begin{split}
        \Omega_\t{ba} = \frac{\hbar G^2}{2 m \Omega_z} \Big( & \frac{\Omega_z-\Omega_k}{(\Omega_z-\Omega_k)^2 + (\kappa/2)^2} \\ 
        & - \frac{\Omega_z+\Omega_k}{(\Omega_z+\Omega_k)^2 + (\kappa/2)^2} \Big),
    \end{split}
\end{equation}
and
\begin{equation}
    \begin{split}
        \gamma_\t{ba} = \frac{\hbar G^2 \kappa}{2 m \Omega_z} \Big( & \frac{1}{(\Omega_z-\Omega_k)^2 + (\kappa/2)^2} \\
        & - \frac{1}{(\Omega_z+\Omega_k)^2 + (\kappa/2)^2} \Big),
    \end{split}
\end{equation}
respectively. 

The stochastic back-action force in \cref{eq:z} is only related to the input field fluctuations,
\begin{equation}
    \hat{F}_\t{ba}[\Omega] = \hbar G \sqrt{\kappa_\t{in}} \left( 
    	\chi_k[\Omega]\hat{a}_\t{in}[\Omega+\Omega_\t{in}] + \t{h.c.} \right)
    %  &- \chi_k^*[\Omega]\hat{a}_\t{in}^\dagger[\Omega-\Omega_\t{in}] \Big),
\end{equation}
and so its statistics is determined by the quantum statistics of the input.
% Notice that we distinguish the fundamental quantum backaction force $\hat{F}_\t{ba}$ from the additional backaction force that arises if the cavity undergoes internal losses, $\hat{F}_\t{ba}^\t{add}$, which would show up as another term in \cref{eq:z} and would be proportional to $\sqrt{\kappa_\t{add}}$.
Assuming that the input field is in a thermal state, the stochastic back-action force spectral density is given
by
\begin{equation}\label{eq:bafull}
\begin{split}
	\bar{S}_{FF}^\t{ba}[\Omega] = \hbar^2 G^2 \kappa_\t{in} 
		\Big[& \abs*{\chi_k[\Omega]}^2 \left( n_\t{th}[|\Omega-\Omega_\t{in}|]+\frac{1}{2} \right) \\ 
		&+ \abs*{\chi_k^*[\Omega]}^2 \left( n_\t{th}[|\Omega+\Omega_\t{in}|]+\frac{1}{2} \right)
		\Big],
\end{split}
\end{equation}
where $n_\t{th}[\Omega] = [\exp(\hbar \Omega/k_B T)-1]^{-1}$ is the average thermal occupation of the 
input field at temperature $T$. Note that in the low temperature limit, $n_\t{th} \rightarrow 0$, but
the stochastic back-action force is still finite, owing to the contribution from the quantum fluctuations
of the input field; this is the unavoidable quantum measurement back-action.

The motion of the electron is conveyed by, and can be inferred from, the field leaking out of the 
cavity \cite{gardiner_input_1985}
\begin{equation}\label{eq:inout}
\begin{split}
    \hat{a}_\t{out}[\Omega] &= \hat{a}_\t{in}[\Omega+\Omega_\t{in}]-\sqrt{\kappa_\t{in}}\hat{a}_k[\Omega] \\
    &= - iG\sqrt{\kappa_\t{in}} \chi_k[\Omega]\, \hat{z}[\Omega] \\
    	&\qquad +(1- \kappa_\t{in} \chi_k[\Omega]) \hat{a}_\t{in}[\Omega+\Omega_\t{in}] \\
    &= -iG\sqrt{\kappa_\t{in}} \chi_k[\Omega] \chi_z^\t{eff}[\Omega]\Big( \hat{F}_\t{dir} +
    \hat{F}_\t{ba} + \hat{F}_\t{imp} \Big),
\end{split}
\end{equation}
where in the last line, the motion $\hat{z}$ is expressed in terms of the forces, and we have defined
the force-equivalent measurement imprecision
\begin{equation}
	\hat{F}_\t{imp}[\Omega] = \frac{i(1-\kappa_\t{in}\chi_k[\Omega])}
	{G\sqrt{\kappa_\t{in}} \chi_k[\Omega]\chi_z^\t{eff}[\Omega]} \hat{a}_\t{in}[\Omega+\Omega_\t{in}],
\end{equation}
which describes the apparent force due to quantum and classical noise from the field used for the measurement.

The force sensitivity of our scheme is clearly limited by the back-action ($\hat{F}_\t{ba}$) 
and imprecision forces ($\hat{F}_\t{imp}$) --- which 
are both fundamentally of a quantum mechanical origin --- and by any extraneous forces that act directly
($\hat{F}_\t{dir}$) on the electron.

In order to realize this force sensitivity in an experiment, we imagine
subjecting the field to homodyne detection (or any other phase-sensitive amplification), 
which will provide access to an arbitrary field quadrature: 
$\hat{q}_\t{out}^\theta[\Omega]=\frac{1}{\sqrt{2}}(\hat{a}_\t{out}[\Omega]e^{-i\theta}
+\hat{a}_\t{out}^\dagger [\Omega]e^{i\theta})$. 
The spectral density of the quadrature of the output field will be of the form,
\begin{equation}\label{eq:Sqqoutimpba}
    \begin{split}
        \bar{S}_{qq}^{\t{out},\theta}[\Omega] \propto & \bar{S}_{FF}^\t{dir}[\Omega] +  \bar{S}_{FF}^\t{ba}[\Omega] +  \bar{S}_{FF}^{\t{imp},\theta}[\Omega]\\
        & + 2 \t{Re}\left[
        \bar{S}_{FF}^{\t{imp,ba},\theta}[\Omega] \right] 
    \end{split}
\end{equation}
The first term represents the direct force noise. The second term is the backaction force noise that arises due to the input field noise being transduced onto the electron\change{, given by \cref{eq:bafull}}, and the third term is the imprecision force noise caused by the same input noise being recycled by the cavity back to the output port. Finally, the last term in \cref{eq:Sqqoutimpba} accounts for possible correlations in the imprecision and backaction forces. The full expressions for these spectral densities are as follows,
\begin{widetext}
    % \begin{equation}\label{eq:bafull}
    %     \bar{S}_{FF}^\t{ba}[\Omega] = \hbar^2 G^2 \kappa_\t{in} \left( \left| \chi_k[\Omega] \right|^2 \left( n_\t{th}[|\Omega-\Omega_\t{in}|]+\tfrac{1}{2} \right) + \left| \chi_k^\dagger[\Omega] \right|^2 \left( n_\t{th}[|\Omega+\Omega_\t{in}|]+\tfrac{1}{2} \right)\right),
    % \end{equation}
    \begin{equation}\label{eq:impfull}
        \bar{S}_{FF}^{\t{imp},\theta}[\Omega] = \frac{1}{G^2 \kappa_\t{in} \left| \chi_z^\t{eff}[\Omega] \right|^2} \frac{\left|1-\chi_k[\Omega]\kappa_\t{in}\right|^2 \left( n_\t{th}[|\Omega-\Omega_\t{in}|]+\frac{1}{2} \right) +\left| 1-\chi_k^*[\Omega]\kappa_\t{in} \right|^2 \left( n_\t{th}[|\Omega+\Omega_\t{in}|]+\frac{1}{2} \right)}{\left| \chi_k[\Omega] e^{-i\theta}-\chi_k^*[\Omega] e^{i\theta} \right|^2},
    \end{equation}
    \begin{equation}\label{eq:impbafull}
        \bar{S}_{FF}^{\t{imp,ba},\theta}[\Omega] = \frac{i \hbar}{\chi_z^\t{eff}[\Omega]} \frac{\left(1-\chi_k[\Omega]\kappa_\t{in}\right)\chi_k[\Omega]^\dagger e^{-i\theta} \left( n_\t{th}[|\Omega-\Omega_\t{in}|]+\frac{1}{2} \right) + \left(1-\chi_k^*[\Omega]\kappa_\t{in}\right)\chi_k^*[\Omega]^\dagger e^{i\theta} \left( n_\t{th}[|\Omega+\Omega_\t{in}|]+\frac{1}{2} \right)}{ \chi_k[\Omega] e^{-i\theta}-\chi_k^*[\Omega] e^{i\theta} }.
    \end{equation}
\end{widetext}

Already from \cref{eq:inout}, it may be surmised that the phase quadrature ($\theta=\pi/2$) conveys information
about the electron's motion. Further, it is necessary to use a high-quality cavity ($\Omega_k \gg \kappa$) to
prolong the interaction between the cavity field and the antenna. Finally, in order to minimize the effect of
dynamical back-action, it is necessary that $\Omega_z - \Omega_k \gg \kappa$. 
Under these conditions, we have the following approximate expressions around $\Omega \sim \Omega_z$,
\begin{align}
    \bar{S}_{FF}^\t{ba}[\Omega] \approx \frac{\hbar^2 G^2 \kappa_\t{in}}
    	{(\Omega-\Omega_k)^{2}+(\kappa/2)^2}  & \left( n_\t{th}[|\Omega-\Omega_\t{in}|] + \frac{1}{2} \right)\label{eq:ba},\\
    \bar{S}_{FF}^{\t{imp},\pi/2}[\Omega]\approx \frac{(\Omega-\Omega_k)^2 +(\kappa/2)^2}{G^2 \kappa_\t{in} \left| \chi_z^\t{eff}[\Omega] \right|^2} & \left( n_\t{th}[|\Omega-\Omega_\t{in}|] + \frac{1}{2} \right) \label{eq:imp},\\
    \bar{S}_{FF}^{\t{imp,ba},\pi/2}[\Omega] \approx \frac{i \hbar}{\chi_z^\t{eff}[\Omega]} \frac{\Omega-\Omega_k}{\Omega+\Omega_k} &\left( n_\t{th}[|\Omega-\Omega_\t{in}|] + \frac{1}{2} \right).
% \end{equation}
\end{align}
On the one hand, measurement imprecision can be reduced by strong coupling (large $G$) between the motion and
the field, however, at the expense of increased back-action. Thus, the continuous measurement scheme outlined
here, in the case of phase quadrature readout, is limited by the fact that
\begin{equation}\label{eq:SQL}
\begin{split}
	\bar{S}_{FF}^\t{dir} + \bar{S}_{FF}^\t{ba} + \bar{S}_{FF}^\t{imp,\pi/2} 
	&\geq \bar{S}_{FF}^\t{dir} + 2\sqrt{\bar{S}_{FF}^\t{ba}\,\bar{S}_{FF}^\t{imp,\pi/2}} \\
	&= \bar{S}_{FF}^\t{dir} + \frac{2\hbar}{\abs{\chi_z^\t{eff}}} 
		\left( n_\t{th} + \frac{1}{2} \right) \\
	&\geq \bar{S}_{FF}^\t{dir} + \frac{\hbar}{\abs{\chi_z^\t{eff}}}.
\end{split}
\end{equation}
The last term in the last line expresses the fundamental limit to the force sensitivity via phase-quadrature
readout in our antenna-mediated coupling scheme --- a form of the Standard Quantum Limit (SQL) for this measurement.

The requirements for high precision continuous force sensing is indicated by \cref{eq:SQL}: 
additional force noises that act directly (i.e. $\hat{F}_\t{dir}$) on the electron will mask the external 
force to be sensed; 
additional readout noises (which can be referred back to an apparent force noise and subsumed in $\hat{F}_\t{imp}$) 
will have the same effect. Further, the limit set by the SQL may itself be large:
indeed, $\hbar/\abs{\chi_z^\t{eff}[\Omega \approx \Omega_z]} \approx \hbar m \gamma_\t{eff} \Omega_z$, so that
extraneous damping of the electron can increase the SQL.

\subsection{Technical limitations}\label{sec:noise}

We now study extraneous noise sources in the above three categories: those which manifest as additional damping 
of the electron motion, those which act as a direct force noise on the electron motion, 
and those which lead to additional readout noise. 

\subsubsection{Additional damping}

Dynamical back-action damping $\gamma_\t{ba}$ accounts for the damping of the electron motion due to
the emission and re-absorption of radiation through the dominant resonant mode of the cavity. 
However, any cavity is still a multi-mode system, so that --- just as in free-space --- the charge is free
to emit Larmor radiation into these modes, but at a rate suppressed by the Purcell factor \cite{Pur46,ParkStr87}.

In addition to dynamical backaction damping, $\gamma_\t{ba}$, the relevant technical damping mechanisms are the cavity-corrected Larmor radiation of the trapped electron, $\gamma_{L}^\t{cav}$, and damping due to losses in the antenna, $\gamma_{A}$,
\begin{equation}\label{eq:gammaz}
    \gamma_\t{eff} = \gamma_{L}^\t{cav} + \gamma_\t{ba} + \gamma_{A}.
\end{equation}
The damping due to dephasing of the three motional degrees of freedom of the electron can be shown to be negligible for the parameter ranges considered here (see \cref{sec:non-idealH}).
The cavity-corrected Larmor damping rate is given by \cite{yokoyama_spontaneous_1995} 
\begin{equation}\label{eq:gammaLz}
    \gamma_L^\t{cav} \approx \gamma_L^\t{free} \left( \frac{(2\pi c \Omega_z^{-1})^3}{V} \right) Q^{-1} = \frac{e^2 \pi^2 }{6 V Q \epsilon_0 m \Omega_z},
\end{equation}
where $m$ is the mass of the electron, $V$ is cavity volume, and  $\epsilon_0$ is the vacuum permittivity, and $Q$ is the cavity quality factor. Note that this expression is approximately valid when the electron is detuned from the cavity resonance and is sufficient as Larmor damping is the subdominant contribution in this regime of operation.
The antenna damping, the dominant damping source, is a result of Ohmic losses of the antenna current \cite{brown_geonium_1986}
\begin{equation}
    \gamma_{A} = \left(\frac{q}{2z_0}\right)^2\frac{R_\t{d}}{m}.
\end{equation}
$R_d$ is given by the antenna geometry and material as $R_d = \rho l/A_a$, where $\rho$ is the resistivity of the material, $l = \pi c /\Omega_z$ is the length of the antenna (defined as half the wavelength of the axial resonant mode), and $A_a$ is the cross-section of the antenna. 

For a general antenna radiating into free space, losses generally occur via dissipative and radiation resistance, and the total resistance is the sum of these two contributions, i.e. $R_\t{d} + R_\t{r}$. However, our case does not involve an antenna in free space. In fact, by embedding the antenna in a microwave cavity resonant with the antenna, the cavity essentially acts as a filter for the radiation, reflecting some radiation back into the antenna. Since the antenna radiation needs to be transmitted by the microwave cavity in order to radiate into free space, the noise associated with the antenna's radiation resistance is essentially the vacuum noise entering the microwave cavity via its output port. Hence, what can be considered radiation resistance in our case will be taken into account at the level of coupling the cavity to the environment.

\subsubsection{Additional direct force noise}

The most important force noise that acts directly on the trapped electron comes from Johnson and dielectric noises 
in the trap electric field, $\bar{S}_{FF}^\t{J} + \bar{S}_{FF}^\t{D}$, and magnetic field fluctuations
(due to the Barkhausen effect), $\bar{S}_{FF}^\t{Bh}$, 
and the two-level-system loss in the cavity, $\bar{S}_{FF}^\t{TLS}$,
\begin{equation}
    \bar{S}_{FF}^\t{dir} = \bar{S}_{FF}^\t{J} + \bar{S}_{FF}^\t{D} + \bar{S}_{FF}^\t{Bh} + \bar{S}_{FF}^\t{TLS}.
\end{equation}
Overall, the motional heating in Penning traps has been shown to be exceptionally low \cite{goodwin_resolved-sideband_2016,borchert_measurement_2019,sawyer_spin_2014,stutter_sideband_2018}.

The most important electric field noise inside a Penning trap comes from the electrode Johnson noise and the electrode surface dielectric losses \cite{kumph_electric-field_2016}. We model the Johnson force noise of the metal electrodes as (similarly to \cite{kumph_electric-field_2016} with a correction factor, see \cref{sec:Efieldnoise})
\begin{equation}\label{eq:S_Fdelta}
    \bar{S}^\t{J}_{FF} [\Omega] (t_m > \delta) \approx 3 q^2
    \begin{cases}
        \frac{k_B T \rho}{2 \pi z^2 \delta[\Omega]} & \text{if } z > \delta, \\
        \frac{k_B T \rho}{2 \pi z^3} & \text{if } z < \delta,
    \end{cases}
\end{equation}
\begin{equation}\label{eq:S_Ftm}
    \bar{S}^\t{J}_{FF} [\Omega] (t_m < \delta) \approx 3 q^2
    \begin{cases}
        \frac{k_B T \rho}{2 \pi z^2 t_m} & \text{if } z > t_m, \\
        \frac{k_B T \rho}{2 \pi z^3} & \text{if } z < t_m,
    \end{cases}
\end{equation}
where we distinguish the cases of the metal thickness $t_m$ being larger or smaller than the skin depth of the metal $\delta$. $\rho$ here is the resistivity of the metal and $z$ is the distance of the charged particle from the metal surfaces of both endcap electrodes. In a similar way, we model the noise due to a thin dielectric layer on the electrode as follows (again with a correction factor compared to \cite{kumph_electric-field_2016})
\begin{equation}\label{eq:S_Edlperp}
    \bar{S}^\t{D}_{FF} [\Omega] = 3 q^2 \frac{3}{4\pi}\frac{\tan\theta}{\epsilon (1 + \tan^2\theta)}\frac{k_B T t_d}{\Omega z^4}.
\end{equation}
Here, $\tan\theta$ and $\epsilon$ are the loss tangent and the permittivity of the dielectric material, respectively, and $t_d$ is the thickness of the dielectric layer.

The dominant effect of magnetic field noise on the electron is the direct force noise in the axial direction due to magnetic fluctuations in the radial plane. We propose to use permanent magnets for the creation of the trap magnetic field\change{, as opposed to superconducting coils that are normally used,} in order to allow for a more compact and mechanically stable setups that minimizes misalignments between the trap and the magnetic field source, and so the noise is expected to be dominated by the Barkhausen effect \cite{barkhausen_zwei_1919}. We model the magnetic force fluctuations as (see \cref{sec:Barkhausen} for the derivation)
\begin{equation}\label{eq:SFFmag}
        \bar{S}^{\t{Bh},\rho}_{FF} [\Omega] = \left| q \bar{\Omega} \bar{\rho} \right|^2 \left( \frac{g_s \mu_0 \mu_B}{2V}\right)^2 \frac{2\alpha}{\Omega^2 + \alpha^2} \frac{T}{T + T_c},
\end{equation}
where $\alpha$ is the highly uncertain magnetic decay constant, and the mean radial coordinate, $\bar{\rho}$, and the mean angular velocity, $\bar{\Omega}$ for the particle trajectory are given by
\begin{equation}\label{eq:rho}
    \bar{\rho} = \sqrt{\frac{2\hbar}{m\sqrt{\Omega_c^2 - 2\Omega_z^2}}}(1+n_\t{th}[\Omega_+] + n_\t{th}[\Omega_-]),
\end{equation}
and
\begin{equation}\label{eq:omegatheta}
    \bar{\Omega} = \frac{\hbar}{m}\frac{1}{\bar{\rho}^2} (n_\t{th}[\Omega_+] - n_\t{th}[\Omega_-]).
\end{equation}
Here $\Omega_c = \frac{q}{m}B_0$, $\Omega_z^2 = \frac{q V_0}{md^2}$, and $\Omega_\pm=\frac{1}{2}(\Omega_c \pm \sqrt{\Omega_c^2 - 2\Omega_z^2})$. Note that the parameters that encode the material properties of the permanent magnet are the spin g-factor $g_s$, the unit cell volume $V$, and the critical temperature $T_c$. We remark that much is still unknown about the Barkhausen effect, and more experimental investigation is required to test and constrain this model.

Another source of magnetic field fluctuations are the thermal fluctuations of charges in the conducting electrodes, which can give rise to Johnson currents \cite{iivanainen_general_2021,lee_calculation_2008,varpula_magnetic_1984,roth_thermal_1998,nenonen_thermal_1996,lamoreaux_feeble_1999,henkel_magnetostatic_2005}. However, our calculations show that this noise is generally far below the other relevant noise sources in this setup and therefore is not discussed here.

Lastly, we consider force noise due to two-level systems (TLS) \cite{anderson_anomalous_1972,jackle_ultrasonic_1972,agarwal_polaronic_2013}, which are the dominant noise source in microwave cavities \cite{gao_physics_2008,oliver_materials_2013}. It is necessary to consider the TLS noise here since our setup requires the presence of dielectric material in the microwave cavity. Note that similarly to the Barkhausen effect, much is still unknown about the TLS effects in microwave resonators. Our model for the TLS-induced force fluctuation is detailed in \cref{sec:TLS} and yields
\begin{equation}
        \bar{S}_{FF}^\t{TLS}[\Omega] = \hbar^2 G^2 \frac{2k_BT}{\mathcal{E}_\t{zp,k}\Omega} \frac{1}{\epsilon_0\epsilon_r\beta}\frac{-\tan\delta_\t{TLS}}{(1-\epsilon_r^{-1})^2+\tan^2\delta_\t{TLS}},
\end{equation}
where $\beta\equiv\int_{V_\t{TLS}} \phi(\textbf{x}) d^3 \textbf{x}$ is the spatial overlap between the TLS volume and the cavity mode function $\phi(\textbf{r})$. There are two processes in amorphous solids with TLSs that contribute to dissipation, namely resonant and relaxation absorption, with the respective loss tangents \cite{enss_tunneling_2005}
\begin{equation}
    \tan\delta_\t{TLS}^\t{res}=\frac{P_0 p^2 \pi}{3\epsilon_0\epsilon_r}\tanh\frac{\hbar\Omega}{2ks_BT},
\end{equation}
and
\begin{equation}
    \tan\delta_\t{TLS}^\t{rel}=\frac{P_0 p^2}{\epsilon_0\epsilon_r}\int^{t_0}_{\tau_\t{min}} \sqrt{1-\frac{\tau_\t{min}}{\tau}}\frac{\Omega}{1+(\Omega\tau)^2}d\tau,
\end{equation}
where $\tau_\t{min}=(A T^3)^{-1}$, and $t_0$ is the timescale of the experiment. TLSs are found to have universal distribution given by $P_0 = 4.35 \times 10^{31} \t{ergs}^{-1} \t{cm}^{-3}$ \cite{jackle_ultrasonic_1972} and $A=10^8 \t{s}^{-1} \t{K}^{-3}$ \cite{enss_tunneling_2005}. The total loss tangent is then the sum of the resonant and relaxation loss tangents (as the dissipated energy adds up),
\begin{equation}
    \tan\delta_\t{TLS}=\tan\delta_\t{TLS}^\t{res}+\tan\delta_\t{TLS}^\t{rel}.
\end{equation}
The parameters $p$ (electric dipole moment of the TLSs) and $\epsilon_r$ can be found in the literature for a variety of TLS host materials.

\subsubsection{Additional imprecision and backaction force noise}

Additional imprecision and backaction force noise comes from internal losses in the walls of the microwave cavity,
\begin{equation}
    \bar{S}_{FF}^\t{read,add} = \bar{S}_{FF}^\t{imp,add} + \bar{S}_{FF}^\t{ba,add}.
\end{equation}
These losses can most generally be modelled through an additional contribution to the total cavity linewidth, $\kappa = \kappa_\t{in}+\kappa_\t{add}$. In much the same way as the thermal bath associated with the input port manifests as the fundamental imprecision-backaction force noise, this additional loss channel will give rise to an aditional backaction and imprecision force noise,
\begin{equation}\label{eq:baadd}
    \bar{S}_{FF}^\t{ba,add}[\Omega] = \hbar^2 G^2 \kappa_\t{add} \left( n_\t{th}[\Omega]+\tfrac{1}{2} \right) \left( \left| \chi_k[\Omega] \right|^2 + \left| \chi_k^*[\Omega] \right|^2\right),
\end{equation}
\begin{equation}\label{eq:impadd}
    \bar{S}_{FF}^\t{imp,add}[\Omega] = \frac{\kappa_\t{add} \left( n_\t{th}[\Omega]+\tfrac{1}{2} \right) }{G^2 \left| \chi_z^\t{eff}[\Omega] \right|^2} \frac{\left|\chi_k[\Omega]\right|^2+\left| \chi_k^\dagger[\Omega] \right|^2}{\left| \chi_k[\Omega]+\chi_k^\dagger[\Omega] \right|^2},
\end{equation}
where we assumed measuring the phase quadrature, $\theta=\pi/2$.

\section{Results}

\begin{figure*}[t]
    \centering
    \includegraphics[width=0.9\linewidth]{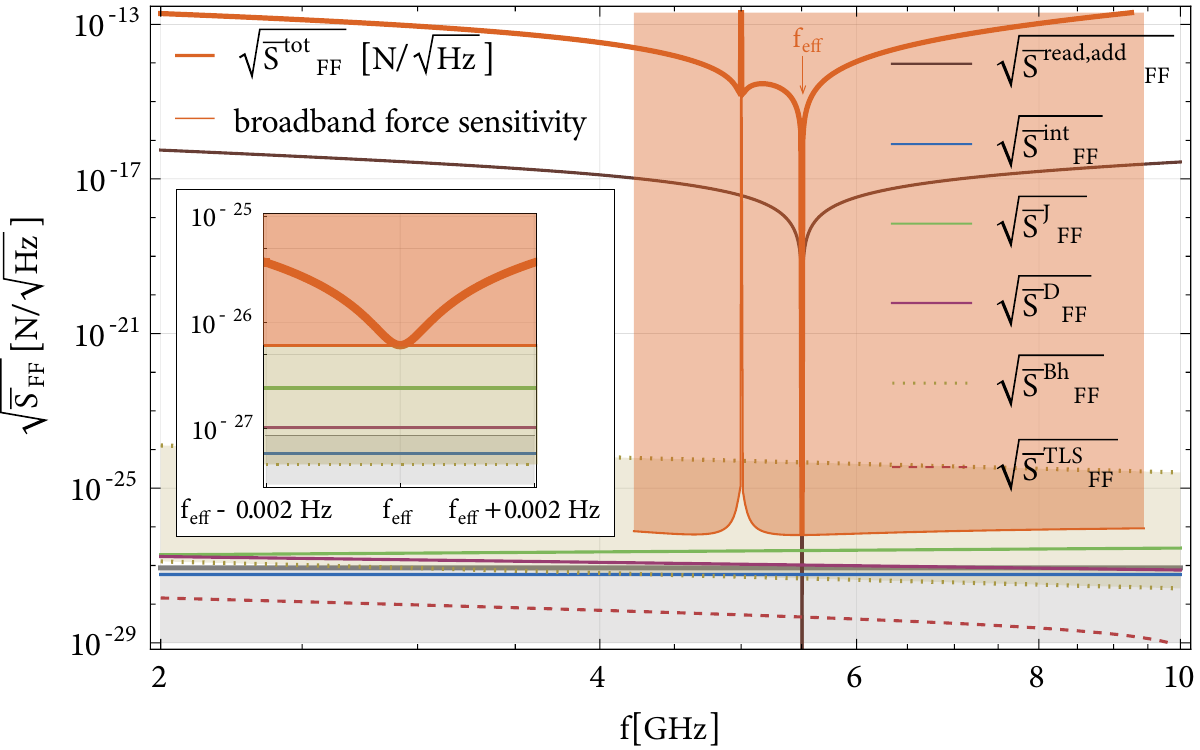} % ~210 words
    \caption{The resultant force noise sensitivity of our proposed setup. The thick orange curve displays the total force noise, $\bar{S}_{FF}^\t{tot}$, for a trap voltage of $19$ V, and the orange shaded region shows the broadband force sensitivity achievable by this detector by tuning the trap voltage between $10$ V and $50$ V. This sensitivity curve is mostly dominated by the imprecision-backaction noise due to losses at the input port of the cavity. The remaining contributions to $\bar{S}_{FF}^\t{tot}$ are: additional imprecision-backaction noise from internal cavity losses $\bar{S}_{FF}^\t{read,add}$ in brown, the intrinsic force noise (including additional damping from dynamical backaction and Ohmic losses in the antenna) $\bar{S}_{FF}^\t{int}$ in blue, the electric Johnson noise from trap electrodes $\bar{S}_{FF}^\t{J}$ in green, the surface-layer dielectric noise from trap electrodes $\bar{S}_{FF}^\t{D}$ in magenta, and the fairly uncertain contributions from Barkhausen noise, $\bar{S}_{FF}^\t{Bh}$, and TLS noise, $\bar{S}_{FF}^\t{TLS}$, in yellow and red, respectively. Note that for $\bar{S}_{FF}^\t{Bh}$, we include a region of plausibility corresponding to a variation of the decay constant between $\alpha/2\pi = 1$ Hz and $\alpha/2\pi = 1$ MHz due to a lack of experimental measurements of this constant. The gray shaded region shows the fundamental limit due to thermal and zero-point motion of a free electron subject to only Larmor damping. The inset shows a zoomed view of the minimum force sensitivity achieved around the electron's effective frequency at $19$ V.}
    \label{fig:force_budget}
\end{figure*}

By considering the scheme depicted in \cref{fig:schematic}a, we have developed a fully quantum theory of such a force transducer and accounted for all relevant quantum, thermal, measurement and technical noises that limit the force sensitivity of this setup. These noises include dielectric fluctuations
in the trap and surface layers in the electrodes \cite{kumph_electric-field_2016}, 
the magnetic Barkhausen noise \cite{barkhausen_zwei_1919}, Johnson noise \cite{Johnson28,Nyquist28} from the trap and antenna, and noise due to fluctuating
two-level systems (TLS) in the cavity walls and intra-cavity dielectrics \cite{gao_physics_2008,oliver_materials_2013}.
As far as the radiation pressure force $\hat{F}_\t{rp}$ is concerned, its fundamental contribution is from
the quantum vacuum fluctuations of the cavity field; however, additional radiation pressure force noise 
from residual thermal noise at the ambient cavity temperature can also be relevant \change{and is modelled here}.

According to our derived interaction Hamiltonian \cref{eq:Hintz}, the motion of the charge ($\hat{z}$) couples to the
amplitude quadrature ($\propto (\hat{a}_k^\dagger +\hat{a}_k)$) of the cavity field; Heisenberg
equations of motion then show that the motion is imprinted onto the phase 
quadrature ($\propto i(\hat{a}_k^\dagger -\hat{a}_k)$) of the cavity field.
Thus measurement of the field leaking out from
the cavity (of damping rate $\kappa$) \cite{gardiner_input_1985}, 
$\hat{a}_\t{out} = \hat{a}_\t{in} - \sqrt{\kappa} \hat{a}_k$, gives access to the continuous motion of the charge.
In particular, the phase quadrature, 
$\hat{q}_\t{out, \pi/2}[\Omega]=\frac{i}{\sqrt{2}}(-\hat{a}_\t{out}[\Omega]+\hat{a}_\t{out}^\dagger [\Omega])$, of the field leaking out can be measured via homodyne detection.
Note that the output quadrature not only depends on the intracavity field fluctuations, which couple to all the aforementioned noise sources, but also on the input field fluctuations, $\hat{a}_\t{in}$. The input field fluctuations that directly leak to the readout quadrature thus contribute sensing noise, 
which sets the imprecision of the field measurement.
Similarly, the thermal noise from the cavity also contributes to the imprecision noise.

In \cref{fig:force_budget}, we plot the resultant force noise sensitivity of our proposed electron force sensor. In what follows, we summarize the choice of parameters that directly produce the noise budget presented in \cref{fig:force_budget}.

We use a single electron (charge $e$, mass $m$) trapped in a $z_0 \approx d=50\ \mu\t{m}$ Penning trap \change{(much smaller than previously realized Penning traps \cite{odom_new_2006,hanneke_new_2008,hanneke_cavity_2011}).} This size enables us to strike a balance between a large enough coupling strength, $G$ (\cref{eq:gz}), and small enough effects of the other noise sources discussed in \cref{sec:noise}. \change{(Note that the resonant frequencies of such a small trap are in the THz regime, far from the GHz frequencies of interest, and so we do not worry about additional motional frequency shifts due to the trap itself \cite{fan_improved_2022}.)} The base trapping potential is chosen to be \change{$V_0=17$} V, which falls into the ballpark range typically used for existing Penning traps. Note that the base voltage affects the axial frequency of the electron ($\propto \sqrt{V_0}$), which therefore provides a useful tuning knob to realize a broadband force detector \change{-- the orange shaded region shows the broadband force sensitivity corresponding to the range $10 {\rm \ V} \leq V_0 \leq 50 {\rm \ V} $}. The value of the trapping magnetic field is chosen to be $B_0=0.5$ T -- this is a sensible estimate of the field that permanent magnet assemblies are able to produce on this scale. The resonant frequency in this case amounts to \change{$f_z \approx 5.5\ \t{GHz}$}.

As for the trap electrodes, we consider the use of gold electrodes ($\rho = 22.1 \text{ n$\Omega$m}$), especially due to its high-conductivity and the low-noise dielectric layer that forms in practice (as compared to materials like copper which are prone to oxidization). We model the trap electrodes as deposited on a dielectric surface (e.g. fused silica), which enables a thickness of \change{$t_m = 5 {\rm \ \mu m}$}. We use the following parameters for the HC film typically formed on gold: $t_d = 2\ \t{nm}$, $\epsilon = 2\epsilon_0$, $\tan\theta = 0.01$ \cite{kumph_electric-field_2016}.

We propose to use samarium cobalt (SmCo) permanent magnets to produce the in-trap magnetic fields. This choice of magnets is informed by previous measurements of magnetic field noise. The critical temperature of SmCo is $T_c=800$ K, and the unit cell properites used here are: $V_\t{uc} = 84.703$ \AA $^3$, $g_s=7.120$. The decay constant $\alpha$ is uncertain as of now, and we display the expected noise for $\alpha/2\pi$ ranging between $1$ Hz and $1$ MHz in \cref{fig:force_budget}.

We assume a symmetric, half-wavelength antenna resonant with the axial frequency, i.e. $l=\pi c/\Omega_z$. The cross-sectional area of the antenna is not tightly bound. We consider a reasonable estimate where one dimension is given by the thickness of the electrodes, $t_m$, and the other one, the antenna width, can be tuned to decrease the imprecision force noise. We have found that an antenna width of \change{$w = 10$} cm helps us obtain good force sensitivity. We also assume the use of gold for the antenna.

\change{In order to estimate the electron-microwave coupling, specifically the term $\xi$ in \cref{eq:gz}, we model the self-capacitances of the trap electrodes and the antenna as $C \approx \epsilon_0 A/t$, where $t$ corresponds to the dimension across which the electric potential is applied, and $A$ corresponds to the cross-sectional area defined by the other two dimensions. In the case of the trap endcap electrode, $C_{\rm electrode} = \epsilon_0 \pi z_0^2/t_m$, and in the case of the antenna $C_{\rm antenna} = \epsilon_0 w t_m /l/2$. }

In light of the discussion surrounding \cref{fig:schematic} c), we choose the resonant frequency of the relevant cavity mode to be \change{$f_k = 5.0\ \t{GHz}$}. A ballpark estimate of the size of the microwave resonator is calculated as ($10\times$ the antenna length)$\times$(the cavity resonant frequency)$\times$($3\times$ the antenna width as the minimum) $\approx (25.6 \rm{\ cm})\times(2.7 \rm{\ cm})\times(15.0 \rm{\ cm})$. Further, we assume an internal quality factor of $Q_\t{int}=10^5$ and a cavity linewidth of $\kappa_\t{add} = \Omega_k/2Q_\t{int}$. We assume a smaller external quality factor, \change{$Q_\t{ext} = 10^2$}, and $\kappa_\t{in} = \Omega_k/2Q_\t{ext}$.

For TLS modelling, we assume that most of the TLS noise comes from the amorphous substrate that hosts the trap + antenna configuration, rather than TLSs on the cavity surfaces. This should be a legitimate assumption, since the intracavity electric field will be larger at the position of this substrate than at the walls of the cavity. Our current designs make use of a fused silica substrate, on which the electrodes and the antenna can be deposited, and so we use the following parameters: $\epsilon_r=3.7$ and $p = 0.5 D$ \cite{golding_intrinsic_1979}. Further, we choose $t_0 = 10^5$ s (although we found that the results are insensitive to the precise choice of this timescale as long as $t_0 \gtrsim 1$ s), and we derive the TLS volume from the antenna length, the trap size, and the antenna width, $V_\t{TLS} = l 2d w$.

Lastly, we drive the cavity on resonance, hence $f_\t{in}=f_k$, and measure the phase quadrature of the outgoing field ($\theta = \pi/2$).

\section{Discussion}

In sum, the ability to continuously monitor the force acting on the electron is impeded by the force
noises that act durectly on it (backaction and other extraneous force), as well as by imprecision noises 
in the measurement of the cavity's output field.
We have accounted for this using a full quantum theory of the measurement process in the Methods section.
The result is that the spectral density of the output quadrature can be expressed in terms of the total 
force noise spectral density,
\begin{equation}\label{eq:SFFtot}
    \begin{split}
        \bar{S}_{qq}^\t{out} \propto \bar{S}_{FF}^\t{tot} & = \bar{S}_{FF}^\t{int} \\ 
        & + \bar{S}_{FF}^\t{imp}+\bar{S}_{FF}^\t{ba}+2\t{Re}[\bar{S}_{FF}^\t{imp,ba}] \\
        & + \bar{S}_{FF}^\t{J} + \bar{S}_{FF}^\t{D} \left( + \bar{S}_{FF}^\t{Bh} + \bar{S}_{FF}^\t{TLS} \right) \\
        & + \bar{S}_{FF}^\t{read,add}.
    \end{split}
\end{equation}
The total force noise, $\bar{S}_{FF}^\t{tot}$, is depicted in \cref{fig:force_budget} for the parameters of our proposed measurement that we elaborate on shortly. The first line of \cref{eq:SFFtot} contains the intrinsic force noise of the electron due to thermal and quantum effects, $\bar{S}_{FF}^\t{int}$
(\cref{eq:SFFint}). The total force sensitivity is further limited
by fundamental readout noises -- imprecision, $\bar{S}_{FF}^\t{imp}$, and backaction, $\bar{S}_{FF}^\t{ba}$ -- due to the
input port of the cavity (second line), a range of technical noises in the third line (electric Johnson noise, $\bar{S}_{FF}^\t{J}$, dielectric noise, $\bar{S}_{FF}^\t{D}$, magnetic Barkhausen noise, $\bar{S}_{FF}^\t{Bh}$, and noise due to two level systems (TLSs), $\bar{S}_{FF}^\t{TLS}$), as well as additional imprecision
and backaction noises due to thermal cavity effects, $\bar{S}_{FF}^\t{read,add} = \bar{S}_{FF}^\t{ba,add} + \bar{S}_{FF}^\t{imp,add}$ (fourth line). Note
that some of these noises, such as the intrinsic or readout noises, are also
affected by additional electron damping due to the dipole interaction and
additional losses. Furthermore, the bracketed terms in the third line of
\cref{eq:SFFtot} are noises whose modeling is highly uncertain, and are hence
excluded from the $\bar{S}_{FF}^\t{tot}$ line in \cref{fig:force_budget}.
Nevertheless, the estimates of these noises are shown in yellow and red in \cref{eq:SFFtot}.

Our proposed transducer utilizes a trapped electron due to its low mass and
thus extremely low intrinsic noise \cref{eq:SFFint}. In addition to the intrinsic noise of this mechanical oscillator, the act of coupling the electron to a microwave cavity as a readout mechanism imposes a fundamental noise floor for this measurement scheme. This noise floor is present even if all extraneous noises are absent and arises due to the tradeoff between the imprecision and backaction forces due to the presence of the microwave cavity: $\bar{S}_{FF}^\t{ba} \propto G^2$, whereas 
$\bar{S}_{FF}^\t{imp} \propto 1/G^2$ (see Methods). Such noise floor is neatly captured in the imprecision-backaction product, also called the Standard Quantum Limit (SQL), which we derive in the Methods section from a fully quantum theory for this new type of coupling,
\begin{equation}
    \bar{S}_{zz}^\t{imp} \bar{S}_{FF}^\t{ba} \approx \hbar^2 \left( n_\t{th}[|\Omega-\Omega_\t{in}|] + \tfrac{1}{2} \right)^2,
\end{equation}
where $n_\t{th}$ is the thermal occupation number of the mechanical oscillator, and $|\Omega-\Omega_\t{in}|$ is the detuning of the oscillator frequency from the frequency of the input field to the cavity.

In practical terms, a decrease in imprecision is thus naively achieved through a smaller cavity volume and a longer antenna (\cref{eq:gz}). However, there are other important technical restrictions that constrain the parameter space. One such restriction comes from factors that affect the electron damping. The damping rate, or conversely the quality factor of the electron oscillator, manifests itself as broadening of the inverted Lorentzian feature in \cref{fig:force_budget}, which is crucial to minimizing the imprecision of the electron motion. We find that the dominant source of damping in our proposed system is Ohmic loss due to image currents in the antenna, which naturally calls for reducing the antenna length and/or extending its cross-sectional area. Furthermore, reducing the cavity volume in general increases the cavity frequency, which subsequently requires an increased axial motional frequency of the electron. This is most conveniently achieved through a smaller trap size, which, however, generally increases the impact of electric and magnetic force noise on the electron. \change{It is for this reason that the most sensitive measurements, such as the electron magnetic moment \cite{odom_new_2006,hanneke_new_2008,hanneke_cavity_2011}, have so far been achieved in centimeter-scale Penning traps. In contrast, the small trapping volume is very crucial to our force-sensitive optomechanical coupling scheme, and hence this proposed setup requires novel technical development before it can be experimentally realized.}

Dynamical backaction damping is another important contribution to the electron damping rate. 
This damping mechanism essentially arises from the retardation in the interaction between the electron's
motion and the cavity field \cite{Gabrielse85}. This effect is largest if the electron motion 
and the cavity are resonant. If the axial frequency of the electron is too close to resonance (i.e. less than the cavity linewidth), the cavity efficiently recycles the antenna radiation and thus amplifies 
its backaction on the electron. In the converse scenario, when the axial frequency is too far away from the cavity frequency (i.e. much more than the cavity linewidth), the antenna radiation into the cavity becomes less efficient and hence imprecision increases. This tradeoff can be very clearly seen in the broadband sensitivity curve in \cref{fig:force_budget}.

In summary, we find a trade-off between the decrease in the imprecision noise and and an increase in the electron damping rate and the effects of other force noises. On this playing field, we have identified a set of detector parameters (described in more detail in the Results section), yielding a micron-scale Penning trap embedded in a cm-scale, relatively low-Q microwave cavity detuned from the electron's frequency by \change{$\sim 0.5\rm{\ GHz}$}, that reduce the total force noise to as low as \change{$6\times 10^{-27} \t{\ N}/ \sqrt{\t{Hz}}$} at \change{$\sim 5.5\rm{\ GHz}$}, as demonstrated in \cref{fig:force_budget}. The coupling strength achieved with these parameters is \change{$G \sim 8\times 10^{10}  \t{\ Hz}/ \t{m}$} (\cref{eq:gz}). We also emphasize that our proposed setup allows us to tune the electron's motional frequency in-situ by adjusting the trap voltage, thus realizing a broadband force detector in the gigahertz regime.

An important component of developing this novel transducer is the consideration of experimental feasibility. The major experimental challenges in realizing this setup and this sensitivity are the fabrication of a microscopic Penning trap, and the lack of a good model for the Barkhausen and TLS losses to guide the design of the magnetic components and the cavity. 
\change{Fabrication techniques for the proposed micron-scale Penning traps are currently being investigated, and
their details are beyond the scope of the current manuscript. Further measurements of Barkhausen and TLS noises in
such small geometries is also necessary to conclusively validate any trap design.}

Accounting for relevant fundamental and technical noise sources, we demonstrate that hybridizing a trapped electron with an antenna and a microwave cavity is expected to enable force detections at the level of \change{$6\times 10^{-27} \t{\ N}/ \sqrt{\t{Hz}}$} in the gigahertz regime. 

\section{Conclusions}

We have proposed a novel broadband force detection scheme using a single trapped electron coupled to a cavity microwave field via an antenna. Despite the size gap between a single electron and a microwave field, efficient coupling can be achieved by using an antenna to amplify the electron's dipole moment\change{, in turn enabling continuous force 
measurement using a trapped electron transducer. Crucially, we exhibit a measurement scheme that can operate at the
Standard Quantum Limit of continuous force sensing, which approaches a sensitivity $\sim 6\times 10^{-27} \t{\ N}/ \sqrt{\t{Hz}}$. This sensitivity improves} on the state-of-the-art by four orders of magnitude and \change{enables} new, exciting experiments at the interface of quantum physics and gravity as well as at the frontier of dark matter search.

\appendix

\section{DAMPING FROM PENNING TRAP NON-IDEALITIES}\label{sec:non-idealH}

An ideal Penning trap of axial and radial half-size $z_0$ and $\rho_0$, respectively, consists of a quadrupole electric potential, $\Phi_0 = \frac{V_0}{2d^2}(z^2-\frac{\rho^2}{2})$, where $d = \sqrt{\frac{1}{2} (z_0^2 + \rho_0^2/2)}$, and a uniform B-field, given by the vector potential $\textbf{A}_0 = \frac{1}{2}B_0 \textbf{e}_\rho \times \textbf{e}_z$, where $\textbf{e}_\rho$ is the radial unit vector in the x-y plane. In such configuration of potentials, the particle motion decomposes into three independent harmonic oscillators: the axial ($\hat{a}_z$), the cyclotron ($\hat{a}_+$) and the magnetron ($\hat{a}_-$) motion \cite{brown_geonium_1986}, i.e.
\begin{equation}\label{eq:H_0_decoupled_main}
    \hat{H}_0 = \hbar\Omega_z \hat{a}^\dagger_z\hat{a}_z + \hbar\Omega_+ \hat{a}^\dagger_+\hat{a}_+ - \hbar\Omega_- \hat{a}^\dagger_-\hat{a}_-.
\end{equation}
Here $\Omega_c = \frac{q}{m}B_0$, $\Omega_z^2 = \frac{q V_0}{md^2}$, $\Omega_l^2 = \Omega_c^2 - 2\Omega_z^2$, and $\Omega_\pm=\frac{1}{2}(\Omega_c \pm \Omega_l)$. The cyclotron frequency, $\Omega_+$ is always the largest frequency scale and thus generally very effectively decoheres through Larmor radiation. 

Real traps inevitably induce corrections to the electric and magnetic trapping fields, either on purpose to enable efficient frequency measurements \cite{ulmer_direct_2011}, or simply due to holes, slits, and manufacturing limitations. In order to calculate the effects that these have on the sensitivity of our proposed force detector, let us start from the generic Hamiltonian of a single charged particle inside an isolated Penning trap
\begin{equation}
    H = \frac{1}{2m}(\textbf{$\Pi$} - q\textbf{A})^2 + q\Phi,
\end{equation}
where $\textbf{A}$ and $\Phi$ are the electromagnetic vector potential and the scalar potential, respectively. Because a Penning trap is a static field configuration, the vector potential is defined as
\begin{equation}
    \nabla \times \textbf{A} = \nabla \times (\textbf{A}_0 + \textbf{A}') = \textbf{B} = B_0 \textbf{e}_z + \textbf{B}'(x,y,z).
\end{equation}
$B_0$ is the (ideal) uniform axial magnetic field, and $\textbf{B}'(x,y,z)$ captures any spatial variations of the magnetic field vector. The vector potential is given by $\textbf{A}_0 = \frac{1}{2}B_0 \textbf{e}_\rho \times \textbf{e}_z$. The scalar potential may be written as
\begin{equation}
    \Phi = \Phi_0 + \Phi'(x,y,z),
\end{equation}
where, again, the first term gives the ideal trapping quadrupole potential $\Phi_0 = \frac{V_0}{2d^2}(z^2-\frac{\rho^2}{2})$, and the second term denotes any non-ideal contributions to the total potential.

Separating the trap fields into their ideal and non-ideal parts, we may write the full Hamiltonian as
\begin{equation}
    H = H_0 + H',
\end{equation}
where $H_0$ is the ideal Penning trap Hamiltonian, and $H'$ captures all trapping non-idealities. More specifically
\begin{equation}\label{eq:H_0}
    H_0 = \frac{1}{2m}\Pi^2 - \frac{q}{m}\textbf{$\Pi$}\cdot \textbf{A}_0 + \frac{q^2}{2m}\textbf{A}_0^2 + q \Phi_0
\end{equation} 
and
\begin{equation}\label{eq:H'}
    H' = - \frac{q}{m}\textbf{$\Pi$}\cdot \textbf{A}' + \frac{q^2}{m}\textbf{A}_0 \cdot \textbf{A}' + \frac{q^2}{2m}\textbf{A}'^2 + q \Phi'.
\end{equation}

Defining $\Omega_c = \frac{q}{m}B_0$, $\Omega_z^2 = \frac{q V_0}{md^2}$, and $\Omega_\pm=\frac{1}{2}(\Omega_c \pm \sqrt{\Omega_c^2 - 2\Omega_z^2})$, and following the quantization procedure and the unitary transformation outlined in \cite{crimin_quantum_2018}, we may rewrite the ideal trap Hamiltonian in \cref{eq:H_0} as
\begin{equation}\label{eq:H_0_decoupled}
    \hat{H}_0 = \hbar\Omega_z \hat{a}^\dagger_z\hat{a}_z + \hbar\Omega_+ \hat{a}^\dagger_+\hat{a}_+ - \hbar\Omega_- \hat{a}^\dagger_-\hat{a}_-.
\end{equation}
Note that we have dropped the constant (vacuum) terms, as they do not contribute to any dynamics in the forthcoming analysis. \cref{eq:H_0_decoupled} indeed shows that the motion of a single charged particle inside a Penning trap can be treated as three independent harmonic oscillators: the axial ($\Omega_z$), the cyclotron ($\Omega_+$) and the magnetron ($\Omega_-$) motions.

\cref{eq:H_0_decoupled} has a much simpler form than \cref{eq:H_0} due to the clever coordinate transformation in \cite{crimin_quantum_2018} which allowed us to understand the physics of the particle motion. By extending this method to the non-ideal Hamiltonian in \cref{eq:H'}, we expect to be able to analyse the effect of these non-idealities on the particle motion from a physically more convenient reference frame. For this method to be valid, we need to associate the coordinate transform used in \cite{crimin_quantum_2018} to a unitary transformation in Hilbert space. It turns out that this coordinate transformation actually corresponds to the well-known beamsplitter unitary $\hat{U} = e^{i\frac{\pi}{4}(\hat{a}^\dagger_x\hat{a}_y + \hat{a}^\dagger_y\hat{a}_x)}$. Hence, we may repeat the aforementioned procedure (i.e. express positions and momenta in terms of $\hat{a}_i$, $\hat{a}^\dagger_i$ and transform to $\hat{a}_\pm$, $\hat{a}^\dagger_\pm$) on $H'$ without any complications.

To the lowest order, the correction terms in \cref{eq:H'} may be written as
\begin{equation}\label{eq:phi'}
    \Phi'(x,y,z) = \frac{\Phi_{40}}{d^4} (x^2 + y^2)^2  + \frac{\Phi_{22}}{d^4} (x^2 + y^2)z^2 + \frac{\Phi_{40}}{d^4} z^4,
\end{equation}
where $d$ is the characteristic trap scale, and considering only the corrections to the axial B-field
\begin{equation}\label{eq:A'}
    \textbf{A}' = \frac{1}{2}\left( \frac{B_{z20}}{d^2} (x^2 + y^2) + \frac{B_{z02}}{d^2} z^2 \right)
    \begin{bmatrix}
        -y \\
        x \\
        0
    \end{bmatrix}.
\end{equation}
Note that it is the quadratic correction to B-field that will be dominant, as this describes the lowest-order curvature of the B-field. A linear contribution would signify just an offset of the maximum axial field from the trap center, which can in principle be minimized. We can see that non-idealities couple all three motions. For convenience, we split the correction Hamiltonian $H'$ into the Hamiltonian due to $\Phi'$, $H_\Phi'$, and a Hamiltonian due to $\textbf{A}'$, $H_A'$, and analyse them sequentially. In both cases, we disregard all constant terms as well as terms which oscillate at a frequency other than the three eigenfrequencies (this is a valid approximation as long as the dominant decay constant for each motion is much smaller than the scale of all motional frequencies).

Realizing that the third term in \cref{eq:H'} will only yield second order terms in the small corrections $B_{z20}$ and $B_{z02}$ that are negligible and skipping tedious algebra, the correction Hamiltonian can be simplified to give
\begin{equation}\label{eq:H'full}
  \begin{split}
      \hat{H}' & = \hbar\Delta\Omega_z \hat{a}^\dagger_z\hat{a}_z + \hbar\Delta\Omega_+\hat{a}^\dagger_+\hat{a}_+ + \hbar\Delta\Omega_-\hat{a}^\dagger_-\hat{a}_- \\
      & + \hbar\Omega_{zz}(\hat{a}^\dagger_z\hat{a}_z)^2 + \hbar\Omega_{++}(\hat{a}^\dagger_+\hat{a}_+)^2 + \hbar\Omega_{--}(\hat{a}^\dagger_-\hat{a}_-)^2 \\
      & + \hbar\Omega_{+z}\hat{a}^\dagger_+\hat{a}_+ \hat{a}^\dagger_z\hat{a}_z + \hbar\Omega_{-z}\hat{a}^\dagger_-\hat{a}_- \hat{a}^\dagger_z\hat{a}_z \\
      & + \hbar\Omega_{+-} \hat{a}^\dagger_+\hat{a}_+\hat{a}^\dagger_-\hat{a}_-.
  \end{split}
\end{equation}
with the frequencies given by
\begin{align} 
    \Delta\Omega_z & = \mathcal{A}(3\Phi_{04} + 2\Phi_{22}) + \mathcal{B} \frac{B_{02}}{2}, \\
    \begin{split}
        \Delta\Omega_+ & = \mathcal{A}(16\Phi_{40} + \Phi_{22}) + 2\mathcal{B} B_{20} \\ 
        & + (\mathcal{B}-\mathcal{C}) \left( 2B_{20} + \frac{B_{02}}{4} \right),
    \end{split}\\
    \begin{split}
        \Delta\Omega_- & = \mathcal{A}(16\Phi_{40} + \Phi_{22}) + 2\mathcal{B} B_{20} \\ 
        & + (\mathcal{B}+\mathcal{C}) \left( 2B_{20} + \frac{B_{02}}{4} \right),
    \end{split}\\
    \Omega_{zz} & = \frac{3}{2}\mathcal{A}\Phi_{04},\\
    \Omega_{++} & = 4\mathcal{A}\Phi_{40} + (\mathcal{B}-\mathcal{C}) B_{20},\\
    \Omega_{--} & = 4\mathcal{A}\Phi_{40} + (\mathcal{B}+\mathcal{C}) B_{20},\\
    \Omega_{+z} & = 2\mathcal{A}\Phi_{22} + \frac{1}{2}(\mathcal{B}-\mathcal{C})B_{02},\\
    \Omega_{-z} & = 2\mathcal{A}\Phi_{22} + \frac{1}{2}(\mathcal{B}+\mathcal{C})B_{02},\\
    \Omega_{+-} & = 16\mathcal{A}\Phi_{40} + 4\mathcal{B} B_{20},
\end{align}
where $\mathcal{A} = \frac{\hbar q}{d^4 m^2 (\Omega_c^2 - 2\Omega_z^2)}$, $\mathcal{B} = \frac{\hbar q^2 B_0}{d^2 m^3 (\Omega_c^2 - 2\Omega_z^2)}$, and $\mathcal{C} = \frac{\hbar q}{d^2 m^2 \sqrt{\Omega_c^2 - 2\Omega_z^2}}$. The first line in \cref{eq:H'full} yields DC motional frequency shifts, the second line contains self-interaction terms (Duffing non-linearities), and finally, the last two lines gives us the coupling terms between all eigenmodes (dephasing terms) that are relevant in the context of decoherence. Note that the symmetry between the cyclotron and magnetron modes has been broken due to the first term in \cref{eq:H'}.

The dephasing terms in \cref{eq:H'full} have the form $\hbar\Omega_{ij}\hat{n}_i\hat{n}_j$ where $i=z,+,-$ and $i\neq j$. The physical effect of such an interaction is the dephasing of energy in the mode $i$ into the mode $j$ and dissipating if mode $j$ decays faster than mode $i$. The corresponding dephasing rate is given by (see \cref{eq:dephasing}),
\begin{equation}
  \delta \gamma_i^j = \frac{\Omega_{ij}^2}{2}\bar{S}_{n_j n_j}[\Omega_i],
\end{equation}
where $\bar{S}_{n_j n_j}$ is the symmetrized spectral density of the number operator of the mode that decays quicker. 
Out of the three motions, the cyclotron mode is the most efficient decay channel, and the magnetron mode has the longest lifetime (i.e. least efficient coupling to the environment). Since it is safe to assume that $\Omega_+ \gg \Omega_z \gg \Omega_- \gg \Omega_{ij}$ for all $i$ and $j$, we can treat the cyclotron mode as a free harmonic oscillator, damped only by Larmor radiation, and the axial mode as subject to a unidirectional dephasing into the cyclotron mode only. We thus consider primarily the dephasing rate of $z$ due to the cylotron mode number operator $\hat{n}_+ = \hat{a}^\dagger_+\hat{a}_+$. As in \cref{eq:Snnapp}, the number fluctuations of the
cyclotron mode are given by,
\begin{equation}\label{eq:Snn}
    \bar{S}_{n_+n_+} [\Omega > 0] = 2n_+(n_+ +1)\frac{\gamma_+}{\gamma_+^2 + (\Omega-2\Omega_+)^2},
\end{equation}
where $n_+$ is the occupation number of the cyclotron mode that can be identified with the Bose occupation number at the cyclotron frequency, $n_\t{th}[\Omega_+]$. This yields the dominant dephasing rate of the axial mode
\begin{equation}\label{eq:gammazplus}
\begin{split}
  \delta\gamma_z = \delta\gamma_z^+ &= n_\t{th}[\Omega_+](n_\t{th}[\Omega_+] +1)\frac{\gamma_+\Omega_{+z}^2}{\gamma_+^2 + (\Omega_z -2\Omega_+)^2} \\
  & \approx \gamma_+ \left( n_\t{th}[\Omega_+] \frac{\Omega_{+z}}{2 \Omega_+}\right)^2, 
\end{split}
\end{equation}
where the last line is valid if $n_\t{th}[\Omega_+]\ll 1$ and is using the fact that the cyclotron motion has a high quality factor (i.e. $\Omega_+ \gg \gamma_+$) even when radiatively damped at the Larmor rate. \Cref{eq:gammazplus} thus provides an additional axial damping term.

In light of the trap requirements presented in the main text, we consider a range of submillimeter-scale Penning traps such that $\Omega_z/2\pi \gtrsim 1$ GHz, ranging from $d = 0.5$ mm ($\Omega_z/2\pi \approx 0.6$ GHz) to $d = 50\ \mu$m ($\Omega_z/2\pi \approx 6$ GHz), all driven by \change{$V_0 = 17$} V and $B_0 = 0.5$ T as in the main text. We find that for fractional non-idealities $\Phi_{22}/V_0$ and $B_{02}/B_0$ in the interval $0.0-1.0$ ($1.0$ meaning that the first order correction is basically the same order of magnitude as the baseline voltage/magnetic field, which is very bad already), the fractional dephasing rate ranges from $10^{-20}$ to $10^{-10}$. Hence, dephasing due to trap non-idealities plays a very negligible part in the overall damping rate $\gamma_z$. This effect becomes even less relevant when we take into account the fact that resistive damping due to the antenna, followed by dynamical backaction damping, is generally going to be the dominant damping mechanism, far more important than Larmor damping considered here as a benchmark. Note that, for the sake of simplicity, $\Phi_{22}/V_0 = B_{02}/B_0$ is assumed, but lifting this constraint does not appreciably change the main result. 

\subsection*{Dephasing rate in the Fock basis}\label{sec:dephasing_rate}

In this section, we derive the dephasing rate of a multi-level, multi-mode harmonic oscillator system subject to a Hamiltonian term $\hat{H}' = \hbar\Omega_{ij}\hat{n}_i\hat{n}_j$, where $i\neq j$ refer to two different modes of the harmonic oscillator. 

Consider a state in the Fock basis $\ket{\Psi} = \sum_{n_j} c_{n_j} \ket*{n_j}$. Unitary evolution dictates
\begin{equation}
    \ket{\Psi(t)} = \hat{U}\ket{\Psi} = e^{-\frac{i}{\hbar}\hat{H}'t}\ket{\Psi} = e^{-i\Omega_{ij}\hat{n}_i\hat{n}_jt}\sum_{n_j} c_{n_j} \ket{n_j}.
\end{equation}
The number operator may be separated into its constant part and the fluctuating part, i.e. $\hat{n}_i = \bar{n}_i + \delta\hat{n}_i$. It is the fluctuating part which induces phase fluctuations to the evolution of $\hat{n}_j$. Hence,
\begin{equation}
    \ket{\Psi(t)} = \sum_{n_j} c_{n_j} e^{-i\Omega_{ij}\delta\hat{n}_i n_jt}\ket{n_j}.
\end{equation}
The evolution prefactor is a stochastic process that causes dephasing of $\ket{\Psi}$. Consider its time-evolved density matrix
\begin{equation}\label{eq:dephasing_rho}
    \hat{\rho} = \langle \ket{\Psi(t)} \bra{\Psi(t)}\rangle = \sum_{n_j}\sum_{m_j} \langle e^{-i\Omega_{ij}\delta\hat{n}_i (n_j-m_j)t} \rangle \ket{n_j}\bra{m_j}.
\end{equation}
We can see that the off-diagonal elements of \cref{eq:dephasing_rho} depend on the time-average of the exponential factor. We may thus identify the dephasing rate $\delta \gamma_i^{j}$ of mode $i$ due to mode $j$ as the time-averaged phase fluctuations with the slowest timescale, i.e.
\begin{equation}
    e^{-\delta \gamma_i^{j}t} \equiv \langle e^{-i\Omega_{ij}\delta\hat{n}_i t}\rangle = \langle e^{-i\phi t}\rangle.
\end{equation}
Following the same procedure as \cite{clerk_introduction_2010}, we assume that the coupling is weak and hence that the phase fluctuations can be regarded statistically independent and Gaussian distributed. From this follows the definition of the dephasing rate $\delta \gamma_i^{j}$ as
\begin{equation}\label{eq:dephasing}
    e^{-\delta \gamma_i^{j}t} = e^{-\frac{\Omega_{ij}^2}{2}\bar{S}_{n_in_i}[\Omega]t} \rightarrow \delta \gamma_i^{j} = \frac{\Omega_{ij}^2}{2}\bar{S}_{n_in_i}[\Omega],
\end{equation}
where $\bar{S}_{n_in_i}$ is the symmetrized spectral density of the number operator fluctuations of the dephasing mode.

\subsection*{Spectral density of the mode number operator}\label{sec:Snn}

Consider the dephasing that follows from \cref{eq:H'full} and again assume that the cyclotron mode decays much faster than the rate of energy exchange between all three modes (as $\Omega_+ \gg \Omega_z \gg \Omega_- \gg \Omega_{ij}$ for all $i$ and $j$). Under such circumstances, the cyclotron mode may be treated as a free harmonic oscillator, damped only by Larmor radiation. The axial mode may then be considered as subject to a unidirectional dephasing into the cyclotron mode only. The magnetron mode, having the longest unperturbed lifetime, is then subject to a unidirectional dephasing into both the axial and the cylotron modes. We may thus write down the following equations of motion
\begin{equation}\label{eq:adotcyclo}
    \dot{\hat{a}}_+ = \left( - i\Omega_+ - \frac{\gamma_+}{2}\right)\hat{a}_+ + \sqrt{\gamma_+}\hat{f}_+,
\end{equation}
\begin{equation}\label{eq:adotz}
    \dot{\hat{a}}_z = \left( - i(\Omega_z+\Omega_{+z}\hat{n}_+) - \frac{\gamma_z}{2}\right)\hat{a}_z + \sqrt{\gamma_z}\hat{f}_z,
\end{equation}
\begin{equation}
    \dot{\hat{a}}_- = \left( - i(\Omega_- +\Omega_{-z}\hat{n}_z +\Omega_{+-}\hat{n}_+) - \frac{\gamma_-}{2}\right)\hat{a}_- + \sqrt{\gamma_-}\hat{f}_-,
\end{equation}
where $\hat{f}_i$ for $i=+,-,z$ are thermal bath operators corresponding to the three motional eigenmodes.

In \cref{eq:adotz}, we see that the dephasing term, $i\Omega_{+z}\hat{n}_+$, is only dependent on the cyclotron motion. Thus, to calculate the dephasing rate of the axial mode, we need to calculate the spectral density of the cyclotron number operator, $\bar{S}_{n_+n_+}$ (\cref{eq:dephasing}). From the Wiener-Khinchine theorem,
\begin{equation}
    \begin{split}
        S_{n_+n_+} [\Omega] & = \int_{-\infty}^{\infty} \langle \hat{a}^\dagger_+(t)\hat{a}_+(t)\hat{a}^\dagger_+(0)\hat{a}_+(0) \rangle e^{-i\Omega t} dt \\
        & = \mathcal{F}[\langle \hat{a}^\dagger_+(t)\hat{a}_+(t)\hat{a}^\dagger_+(0)\hat{a}_+(0) \rangle],
    \end{split}
\end{equation}
where $\mathcal{F}$ denotes Fourier transform. This equation can be simplified by applying Wick's theorem as follows
\begin{equation}
    \begin{split}
        S_{n_+n_+} [\Omega] & = \mathcal{F}[\langle \hat{a}^\dagger_+(t)\hat{a}_+(t)\rangle \langle\hat{a}^\dagger_+(0)\hat{a}_+(0) \rangle] \\
        &  + \mathcal{F}[\langle \hat{a}^\dagger_+(t)\hat{a}_+(0)\rangle \langle\hat{a}_+(t)\hat{a}^\dagger_+(0) \rangle] \\
        & = \mathcal{F}[\langle \hat{a}^\dagger_+(0)\hat{a}_+(0)\rangle^2] + S_{a^\dagger_+ a_+} \star S_{a_+ a^\dagger_+}.
    \end{split}
\end{equation}
The second term of the second equality follows by the convolution theorem. The first term is proportional to $n_+^2 \mathcal{F}[1] \propto \delta [\Omega]$. Since we are ultimately interested only in the positive frequencies of the spectral density, this term does not contribute.
From \cref{eq:adotcyclo}, we have that
\begin{equation}
    S_{a^\dagger_+ a_+} [\Omega] = |\chi_+[\Omega]|^2 S_{f^\dagger_+ f_+} 
\end{equation}
and 
\begin{equation}
    S_{a_+ a_+^\dagger} [\Omega] = |\chi_+[\Omega]|^2 S_{f_+ f^\dagger_+} 
\end{equation}
with 
\begin{equation}
    \chi_+[\Omega] = \frac{\sqrt{\gamma_+}}{-i(\Omega-\Omega_+) + \frac{\gamma_+}{2}}.
\end{equation}
We also have for a thermal bath at temperature $T$ that $S_{f^\dagger_+ f_+} = n_+ = S_{f_+ f^\dagger_+} - 1 $. Convolving the two spectral densities as
\begin{equation}
    \bar{S}_{n_+n_+} [\Omega > 0] = n_+(n_+ +1) \int |\chi_+[\Omega']|^2 |\chi_+[\Omega-\Omega']|^2 \frac{d\Omega'}{2\pi}
\end{equation}
we obtain
\begin{equation}\label{eq:Snnapp}
    \bar{S}_{n_+n_+} [\Omega > 0] = 2n_+(n_+ +1)\frac{\gamma_+}{\gamma_+^2 + (\Omega-2\Omega_+)^2}
\end{equation}
as used in \cref{eq:Snn}.

\section{ELECTRIC FIELD NOISE MODEL}\label{sec:Efieldnoise}

In a Penning trap, the static electric field used to trap the particle is created using an assembly of electrodes, each of which are typically deposited on a dielectric substrate. Fluctuations of this electric field cause motional heating which leads to decoherence of the trapped particle's motional state. Motional heating is studied extensively among the ion trap community (e.g. \cite{brownnutt_ion-trap_2015}), however, most electric field noise measurements to-date come from experiments in Paul traps, rather than Penning traps (the only measurements of Penning trap heating rates that we are aware of are \cite{goodwin_resolved-sideband_2016,borchert_measurement_2019} for single ions; \cite{sawyer_spin_2014,stutter_sideband_2018} for ion crystals). 

It has recently been shown that the measured electric field noise agrees well with the noise expected due to the imaginary (loss) part of the dielectric constant of the surrounding dielectric material abundant in many Paul traps \cite{teller_heating_2021}. In light of this recent result, we suspect that the dominant electric field noise mechanism in Paul traps, i.e.  the dielectric loss, is suppressed in Penning traps which typically provide minimum exposure of the trapped particle to dielectric surfaces. We therefore expect that the electric field force noise in our setup comes in two forms: Ohmic/Johnson losses in the conductor, and losses due to dielectric thin films on the exposed electrode surfaces. Both of these noise sources are well modelled in \cite{kumph_electric-field_2016} (and \cite{henkel_loss_1999}), and we cite their results in this section. Note that these authors considered a single charge above a flat metal surface, but in our case, we are concerned with a charge inside a cylindrical trap. Here we apply their results directly as an estimate of the electric-field noise.

We model the Johnson force noise of the metal electrodes as
\begin{equation}\label{eq:S_Fdelta}
    \bar{S}^\t{J}_{FF} [\Omega] (t_m > \delta) \approx 3 q^2
    \begin{cases}
        \frac{k_B T \rho}{2 \pi z^2 \delta[\Omega]} & \text{if } z > \delta, \\
        \frac{k_B T \rho}{2 \pi z^3} & \text{if } z < \delta,
    \end{cases}
\end{equation}
\begin{equation}\label{eq:S_Ftm}
    \bar{S}^\t{J}_{FF} [\Omega] (t_m < \delta) \approx 3 q^2
    \begin{cases}
        \frac{k_B T \rho}{2 \pi z^2 t_m} & \text{if } z > t_m, \\
        \frac{k_B T \rho}{2 \pi z^3} & \text{if } z < t_m,
    \end{cases}
\end{equation}
where we distinguish the cases of the metal thickness $t_m$ being larger or smaller than the skin depth of the metal $\delta$. Here, we have added a factor of 3 with respect to the result in \cite{kumph_electric-field_2016} as inside the trap, the axial force comes from the perpendicular noise from two endcap electrodes and the parallel noise from the ring electrode (and the parallel field noise is half the perpendicular noise \cite{kumph_electric-field_2016}). Additionally, $\rho$ here is the resistivity of the metal and $z$ is the distance of the charged particle from the metal surfaces of both endcap electrodes.

In a similar way, we model the noise due to a thin dielectric layer on the electrode as follows \cite{kumph_electric-field_2016}
\begin{equation}\label{eq:S_Edlperp}
    \bar{S}^\t{D}_{FF} [\Omega] = 3 q^2 \frac{3}{4\pi}\frac{\tan\theta}{\epsilon (1 + \tan^2\theta)}\frac{k_B T t_d}{\Omega z^4}.
\end{equation}
Again, the factor of 3 is incorporated for the same reason as in the Johnson noise case. Here, $\tan\theta$ and $\epsilon$ are the loss tangent and the permittivity of the dielectric material, respectively, and $t_d$ is the thickness of the dielectric layer. 

Note that the spectrum of the Johnson noise is not always white. The skin depth generally gets smaller for higher probe frequencies (but still well below the plasma frequency), which makes the noise spectrum an increasing function of the frequency once the skin depth is smaller than both the thickness of the metal and the distance to the trapped particle. This increasing trend breaks down at higher frequencies where the model is no longer valid \cite{kumph_electric-field_2016}. While increasing the thickness of the electrode suppresses the Johnson noise acting on the particle, it also decreases the cutoff frequency above which the noise begins to grow.

\section{BARKHAUSEN NOISE MODEL}\label{sec:Barkhausen}

The axial magnetic field required for Penning traps can generally be produced in two ways, namely using superconducting coils, and permanent magnets, both of which generate magnetic field fluctuations. In the case of superconducting coils, the fluctuations of the input current will manifest as fluctuations in the generated magnetic field. In the latter case, the spin domains that make up all permanent magnets will be subject to thermal fluctuations from the environment, which can cause individual spins or whole spin domains to rotate or flip. This stochastic rotations of spins/spin domains in a permanent magnet is known as the Barkhausen effect \cite{barkhausen_zwei_1919}. In this analysis we focus on magnetic field noise of the Barkhausen origin, as the use of permanent magnets allows for more compact and mechanically stable setups that minimize misalignments between the trap and the magnetic field source. 

The simplest description of permanent magnets to zeroth order is given by the Weiss mean field theory (MFT). By definition, the MFT smooths over any variance in magnetisation, and so it is not possible to obtain any prediction from the MFT directly. However, we may derive the lowest order perturbation around the MFT by using the MFT theory Hamiltonian
\begin{equation}
    \begin{split}
        H_\mathrm{MFT} = & -\frac{1}{2} \sum_{<i,j>} J \textbf{s}_i \cdot \textbf{s}_j + g_s\mu_B \textbf{B} \cdot \sum_i \textbf{s}_i \\
        = & \sum_i g_s\mu_B \langle \textbf{B}_\mathrm{eff} \rangle \cdot \textbf{s}_i,
    \end{split}
\end{equation}
where $Jz = 4 k_B T_c$ and $\langle \textbf{B}_\mathrm{eff} \rangle = \textbf{B} - \frac{4 k_B T_c}{g_s\mu_B}\langle \textbf{s} \rangle$, and expanding the spin vector as $\textbf{s}_i = \langle \textbf{s} \rangle + \delta \textbf{s}_i$. Note that we have dropped the index on the mean spin, as the MFT imposes that all sites have the same mean spin. This yields the correction Hamiltonian
\begin{equation}
    \delta H_\mathrm{MFT} = \sum_i g_s\mu_B \langle \textbf{B}_\mathrm{eff} \rangle \cdot \delta\textbf{s}_i.
\end{equation}
As an upper bound on the magnetisation fluctuations we expect experimentally, we consider the spins to be constrained to only one dimension. The change in spin can therefore be $\pm \frac{1}{2}$. The variance in magnetisation is related to the magnetic susceptibility as
\begin{equation}
    \chi_m = \mu_0 \lim_{B\rightarrow 0} \frac{\partial M}{\partial B} = \frac{\mu_0 V}{k_B T} \t{Var}(M)
\end{equation}
where magnetisation is defined as $\langle M \rangle = - \frac{g_s \mu_B}{V}\langle s \rangle$. In this definition, $g_s \mu_B$ is the magnetic moment of the unit cell of the magnetic material and $V$ is the unit cell volume. Using the usual thermodynamic relations
\begin{equation}
    Var(M) \equiv \langle M^2 \rangle - \langle M \rangle ^2 = - k_B T V^{-2} \lim_{B \rightarrow 0} \left( \frac{\partial^2 F}{\partial B^2}\right)_T
\end{equation}
we arrive at the following equation for the fluctuations around the mean magnetisation
\begin{equation}\label{eq:varM}
    \t{Var}(M) = \left( \frac{g_s \mu_B}{2V}\right)^2 \frac{1}{\frac{T_c}{T} + \cosh^2 \left( \frac{2 V \langle M \rangle}{g_s \mu_B} \frac{T_c}{T} \right)}
\end{equation}
\Cref{eq:varM} thus represents the variance in the magnetisation that is driven by thermally induced spin flips.

In order to obtain the frequency distribution of the variance in magnetisation, we write down the most general equation of motion for the magnetization,
\begin{equation}\label{eq:Mdotgen}
    \dot{\textbf{M}} = -\alpha \textbf{M} - \lambda \textbf{M}\times\textbf{H} - \mu \textbf{M}\times (\textbf{M}\times\textbf{H}),
\end{equation}
where $\alpha$, $\lambda$, and $\mu$ are time independent constants, and $\textbf{H}$ is the external magnetic field. This equation holds because we have expanded the vector $\dot{\textbf{M}}$ in terms of three linearly independent vectors in 3D.

\Cref{eq:Mdotgen} can be related to existing models of magnetization evolution, namely the Landay-Lifshitz model and the Gilbert model. The Landau-Lifshitz model phenomenologically assumes that $\alpha=0$ and $\mu=\lambda=\lambda_L$,
\begin{equation}\label{eq:LL}
    \dot{\textbf{M}} = - \lambda_L \textbf{M}\times\textbf{H} - \lambda_L \textbf{M}\times (\textbf{M}\times\textbf{H}).
\end{equation}
This implies that the magnetization magnitude, $M = |\textbf{B}|$, is a constant of motion,
\begin{equation}
    \frac{d(M^2)}{dt} = \frac{d}{dt}(\textbf{M}\cdot\textbf{M})=\frac{1}{2}\textbf{M}\cdot\dot{\textbf{M}} = 0.
\end{equation}
In the Gilbert model, the equation of motion reads 
\begin{equation}\label{eq:Gilbert}
    \dot{\textbf{M}} = - \lambda_G \textbf{M}\times\textbf{H} - \lambda_G \eta \textbf{M}\times\dot{\textbf{M}}.
\end{equation}
Substituting the LHS of \cref{eq:Gilbert} into the $\dot{\textbf{M}}$ on the RHS of the same equation, we show that 
\begin{equation}
    \begin{split}
        \dot{\textbf{M}} & = - \lambda_G \textbf{M}\times\textbf{H} \\
        & - \lambda_G \eta \textbf{M}\times \left( - \lambda_G \textbf{M}\times\textbf{H} - \lambda_G \eta \textbf{M}\times\dot{\textbf{M}} \right)\\
        & = - \lambda_G \textbf{M}\times\textbf{H} + \lambda_G^2 \eta \textbf{M}\times \textbf{M}\times\textbf{H} \\
        & + (\lambda_G \eta)^2 \left( (\textbf{M}\cdot\dot{\textbf{M}})\textbf{M} - M^2 \dot{\textbf{M}} \right) \\
        & = - \lambda_G \textbf{M}\times\textbf{H} + \lambda_G^2 \eta \textbf{M}\times \textbf{M}\times\textbf{H} - (\lambda_G \eta M)^2 \dot{\textbf{M}}
    \end{split}
\end{equation}
Hence, 
\begin{equation}
    \begin{split}
        \dot{\textbf{M}} = & - \left( \frac{\lambda_G}{1+(\lambda_G \eta M)^2} \right) \textbf{M}\times\textbf{H} \\
        & - \left( \frac{- \lambda_G^2 \eta}{1+(\lambda_G \eta M)^2} \right) \textbf{M}\times (\textbf{M}\times\textbf{H}),
    \end{split}
\end{equation}
which is consistent with the Landau-Lifshitz model in \cref{eq:LL} and the prediction that the magnetization is a constant of motion under the influence of an external magnetic field.

In order to model any fluctuation-dissipation dynamics, we need to extend these models by regarding $\alpha\neq 0$, since we know from experiments that it is possible to impulsively excite the magnetization and observe its damped relaxation (i.e. Barkhausen effect). Hence,
\begin{equation}\label{eq:dM2}
    \frac{d(M^2)}{dt} = -\frac{\alpha}{2}M^2.
\end{equation}
Here, the sign of $\alpha$ determines the magnetization at long times. Since the magnetization has to be finite, $\alpha$ cannot be negative. If $\alpha>0$, \cref{eq:dM2} suggests that the magnet is demagnetized eventually, in contradiction with what one expects in thermal equilibrium below the Curie temperature.

The remedy to this apparent contradiction is a stochastic Langevin force, $\textbf{N}$,
\begin{equation}\label{eq:MdotLan}
    \dot{\textbf{M}} = -\alpha \textbf{M} - \lambda \textbf{M}\times\textbf{H} - \mu \textbf{M}\times (\textbf{M}\times\textbf{H}) + \textbf{N}.
\end{equation}
In the case of zero external field, \cref{eq:MdotLan} simplifies to
\begin{equation}\label{eq:deltadotM}
    \dot{\textbf{M}} = - \alpha \textbf{M} + \textbf{N},
\end{equation}
where $\alpha$ can be now interpreted as a decay constant associated with magnetisation relaxation. In terms of the symmetrized spectral densities
\begin{equation}
    \bar{S}_{MM} [\Omega] = \frac{\bar{S}_{NN} [\Omega]}{\Omega^2 + \alpha^2}.
\end{equation}
We can now express $\bar{S}_{NN}$ by Fourier transforming \cref{eq:deltadotM} and integrating over the corresponding spectral density $\t{Var}(M) \equiv \int \bar{S}_{MM} \frac{d\Omega}{2\pi}$. We obtain 
\begin{equation}
    \bar{S}_{NN} [\Omega] = \frac{\t{Var}(M)}{\int_{-\infty}^{\infty} \frac{1}{\Omega^2 + \alpha^2}\frac{d\Omega}{2\pi}} = 2\alpha \t{Var}(M).
\end{equation}
Hence, the magnetisation spectral density we expect for a 1D ferromagnetic spin system is 
\begin{equation}\label{eq:SMMfull}
    \bar{S}_{MM} [\Omega] = \frac{2\alpha}{\Omega^2 + \alpha^2} \left( \frac{g_s \mu_B}{2V}\right)^2 \frac{1}{\frac{T_c}{T} + \cosh^2 \left( \frac{2 V \langle M \rangle}{g_s \mu_B} \frac{T_c}{T} \right)}.
\end{equation}

Since not much is known about the Barkhausen effect in hard magnets, our model can only serve as an estimate at best. In order to get the upper bound on the magnetisation fluctuations we expect, we set $\langle M \rangle = 0$ according to \cref{eq:SMMfull}. Hence, the radial magnetic field noise due to thermal fluctuations in the permanent magnet magnetisation is given by 
\begin{equation}\label{eq:SBBmag}
    \begin{split}
        \bar{S}^{\t{Bh},\rho}_{BB} [\Omega] = & \mu_0^2 \bar{S}^\rho_{MM} [\Omega] \\
        = & \left( \frac{g_s \mu_0 \mu_B}{2V}\right)^2 \frac{2\alpha}{\Omega^2 + \alpha^2} \frac{T}{T + T_c}.
    \end{split}
\end{equation}
The parameters that encode the material properties of the permanent magnet are the spin g-factor $g_s$, the unit cell volume $V$, and the critical temperature $T_c$.

We now need to relate the radial magnetic field fluctuations to the axial magnetic force scting on the trapped electron. Assuming no angular magnetic field fluctuations due to the symmetry of the setup, the axial Lorentz force can be written as
\begin{equation}
    F_z^\t{B} = q v_\theta B_\rho = q \Omega \rho B_\rho,
\end{equation}
where $\Omega$ and $\rho$ correspond to the angular velocity and the radial coordinate of the trapped particle trajectory (which is itself fluctuating). Since the mean radial magnetic field is zero, $B_\rho = \delta B_\rho$ and we may write down the following relation for the axial magnetic force fluctuations 
\begin{equation}
    \delta F_z^\t{B} = q \bar{\Omega} \bar{\rho} \delta B_\rho.
\end{equation}
Transducing the magnetic field fluctuations to a force noise thus boils down to calculating the mean radial coordinate, $\bar{\rho}$, and the mean angular velocity, $\bar{\Omega}$ for the particle trajectory.

Using the transformations outlined in \cite{crimin_quantum_2018}, we can write $\bar{\rho}$ and $\bar{\Omega}$ in terms of the number operators corresponding to the three harmonic motions of the particle. Note that we use $\rho \equiv \sqrt{x^2 + y^2} = \bar{\rho} + \delta \rho$ and $\Omega \equiv \frac{d\theta}{d t} = \frac{d }{d t} \arctan{y/x} = \bar{\Omega} + \delta \Omega$. This trivial calculation yields
\begin{equation}\label{eq:rho}
    \bar{\rho} = \sqrt{\frac{2\hbar}{m\omega_l}}(1+\bar{n}_+ + \bar{n}_-)
\end{equation}
and
\begin{equation}\label{eq:omegatheta}
    \bar{\Omega} = \frac{\hbar}{m}\frac{1}{\bar{\rho}^2} (\bar{n}_+ - \bar{n}_-).
\end{equation}

Using \cref{eq:rho,eq:omegatheta,eq:SBBmag} and dropping the superscript $\rho$, we can now write down the axial force noise spectral density that we expect based on the spectral density of the radial magnetic field fluctuations as
\begin{equation}\label{eq:BtoF}
    \begin{split}
        \bar{S}^{\t{Bh}}_{FF} [\Omega] & = |q \bar{\Omega} \bar{\rho} |^2  \bar{S}^{\t{Bh}}_{BB} [\Omega]\\
        & =  \left| q \bar{\Omega} \bar{\rho} \right|^2 \left( \frac{g_s \mu_0 \mu_B}{2V}\right)^2 \frac{2\alpha}{\Omega^2 + \alpha^2} \frac{T}{T + T_c}.
    \end{split}
\end{equation}

\section{TWO-LEVEL SYSTEM (TLS) NOISE MODEL}\label{sec:TLS}

Two-level systems (TLS) are a class of dissipating systems in amorphous materials \cite{anderson_anomalous_1972,jackle_ultrasonic_1972,agarwal_polaronic_2013}, especially prevalent in superconducting qubit research \cite{gao_physics_2008,oliver_materials_2013}. They can be modelled as a particle in a double potential well, where the two spatial states correspond to the two potential minima, and tunnelling between the two eigenstates is allowed. If the TLS system is charged, its tunnelling properties create a dipole moment that interacts with an external electric field. Therefore, amorphous materials that exhibit TLS behaviour essentially form a bath of two-level oscillators that can couple to the electric field and thus dissipate energy into uncontrolled degrees of freedom.

We model the intracavity electric field fluctuations due to the TLSs via the TLS loss tangent, $\tan\delta_\t{TLS}$, as described by Enss and Hunklinger \cite{enss_tunneling_2005}. To link the loss tangent of the TLS material to the spectral density of the electric field fluctuations, we apply the fluctuation-dissipation theorem \cite{kubo_fluctuation-dissipation_1966}. We assume a dipole coupling $\textbf{p}_\t{TLS} \cdot \textbf{E}_k$ between the collection of TLS dipole moments $\textbf{p}_\t{TLS}$ and the intracavity microwave field $\textbf{E}_k$. We make two assumptions: the material is a linear, homogeneous and isotropic medium, and we only consider the 1D version of this problem. Hence,
\begin{equation}\label{eq:p-E}
    p_\t{TLS}=\int_{V_\t{TLS}} P dV = \epsilon_0 (\epsilon_r-1) \int_{V_\t{TLS}} E_k dV,
\end{equation}
where $P$ is the polarisation within the TLS medium, $\epsilon_r-1$ is the electric susceptibility (commonly labelled as $\chi_r$; we do not employ this notation so as not to confuse it with the susceptibility relevant for the fluctuation-dissipation theorem), and the integrals are over the volume of the TLS material. Inside the cavity, the electric field can be relabelled as 
\begin{equation}\label{eq:Emode}
    E_k(\textbf{x},t)\rightarrow\phi(\textbf{x})E_k(t),
\end{equation}
where the spatial mode part is defined such that $\int \phi(\textbf{x}) d^3\textbf{x} = 1$ (the integral is over all space), and $E_k(t)$ is the time-dependent electric field whose fluctuations we're interested in. We can thus use \cref{eq:p-E,eq:Emode} to derive the susceptibility
\begin{equation}\label{eq:chiTLS}
    \chi_\t{pp}^\t{TLS} = \epsilon_0 (\epsilon_r-1) \int_{V_\t{TLS}} \phi(\textbf{x}) d^3 \textbf{x}.
\end{equation}
Furthermore, we can rewrite the relative permittivity in terms of the loss angle, defined as $\epsilon_r\rightarrow\epsilon_r(1+i\tan\delta_\t{TLS})$. The fluctuation-dissipation theorem \cite{kubo_fluctuation-dissipation_1966} thus reads 
\begin{equation}
    \begin{split}
        \bar{S}_{EE}^\t{TLS}[\Omega] & =\frac{2k_BT}{\Omega} \t{Im} \left[ \chi_\t{pp}^{\t{TLS},-1} [\Omega] \right]\\
        & = \frac{2k_BT}{\Omega} \frac{1}{\epsilon_0\epsilon_r\beta}\frac{-\tan\delta_\t{TLS}}{(1-\epsilon_r^{-1})^2+\tan^2\delta_\t{TLS}},
    \end{split}
\end{equation}
where $\beta\equiv\int_{V_\t{TLS}} \phi(\textbf{x}) d^3 \textbf{x}$ is the spatial overlap between the TLS volume and the electric field mode function $\phi(\textbf{x})$ (if the linear dimensions of the TLS volume are much smaller than the electric field wavelength then this just becomes the total TLS volume). The intracavity field fluctuations are thus determined by the TLS loss angle.

The TLS loss angle is derived by Enss and Hunklinger \cite{enss_tunneling_2005} who consider two processes in amorphous solids with TLSs that contribute to dissipation: resonant absorption and relaxation absorption. The respective loss tangents are given by \cite{enss_tunneling_2005}
\begin{equation}
    \tan\delta_\t{TLS}^\t{res}=\frac{P_0 p^2 \pi}{3\epsilon_0\epsilon_r}\tanh\frac{\hbar\Omega}{2ks_BT},
\end{equation}
and
\begin{equation}
    \tan\delta_\t{TLS}^\t{rel}=\frac{P_0 p^2}{\epsilon_0\epsilon_r}\int^{t_0}_{\tau_\t{min}} \sqrt{1-\frac{\tau_\t{min}}{\tau}}\frac{\Omega}{1+(\Omega\tau)^2}d\tau,
\end{equation}
where $\tau_\t{min}=(A T^3)^{-1}$, and $t_0$ is the timescale of the experiment. TLSs are found to have universal distribution with the relevant parameters given by $P_0 = 4.35 \times 10^{31} \t{ergs}^{-1} \t{cm}^{-3}$ \cite{jackle_ultrasonic_1972} and $A=10^8 \t{s}^{-1} \t{K}^{-3}$ \cite{enss_tunneling_2005}. The total loss tangent is then the sum of the resonant and relaxation loss tangents,
\begin{equation}
    \tan\delta_\t{TLS}=\tan\delta_\t{TLS}^\t{res}+\tan\delta_\t{TLS}^\t{rel}.
\end{equation}
The parameters $p$ (electric dipole moment of the TLSs) and $\epsilon_r$ can be found in the literature for a variety of TLS host materials.

These intracavity electric field fluctuations are converted to an effective force noise on the electron via the coupling constant,
\begin{equation}
    \begin{split}
        \bar{S}_{FF}^\t{TLS}[\Omega] & = \left| \hbar \frac{G}{\mathcal{E}_\t{zp,k}}\right|^2  \bar{S}_{EE}^\t{TLS}[\Omega]\\
        & = \hbar^2 G^2 \frac{2k_BT}{\mathcal{E}_\t{zp,k}\Omega} \frac{1}{\epsilon_0\epsilon_r\beta}\frac{-\tan\delta_\t{TLS}}{(1-\epsilon_r^{-1})^2+\tan^2\delta_\t{TLS}},
    \end{split}
\end{equation}
as used in the main text.

\bibliography{refs_new}

\end{document}